\documentclass[fleqn, usenatbib]{mnras}

\usepackage{newtxtext,newtxmath}

\usepackage[T1]{fontenc}

\DeclareRobustCommand{\VAN}[3]{#2}
\let\VANthebibliography\thebibliography
\def\thebibliography{\DeclareRobustCommand{\VAN}[3]{##3}\VANthebibliography}

\usepackage{xcolor}

\usepackage{graphicx}	
\usepackage{amsmath}	

\usepackage{listings}
\usepackage{color}
\definecolor{dkgreen}{rgb}{0,0.6,0}
\definecolor{gray}{rgb}{0.5,0.5,0.5}
\definecolor{mauve}{rgb}{0.58,0,0.82}
\definecolor{golden}{rgb}{0.86,0.65,0.01}
\defcitealias{Rappaport1983}{RVJ}

\title[Gaia BH2]{A red giant orbiting a black hole}

\author[El-Badry et al.]{Kareem El-Badry$^{1,2,3,4}$\thanks{E-mail: kareem.el-badry@cfa.harvard.edu}, Hans-Walter Rix$^{3}$, Yvette Cendes$^1$,  Antonio C. Rodriguez$^4$, \newauthor Charlie Conroy$^1$, Eliot Quataert$^5$, Keith Hawkins$^6$, Eleonora Zari$^3$, Melissa Hobson$^3$, Katelyn Breivik$^7$, \newauthor Arne Rau$^8$,  Edo Berger$^1$, Sahar Shahaf$^{9}$,  Rhys Seeburger$^{3}$, Kevin B. Burdge$^{10}$,  David W. Latham$^{1}$, \newauthor Lars A. Buchhave$^{11}$, Allyson Bieryla$^{1}$, Dolev Bashi$^{12}$,  Tsevi Mazeh$^{12}$, Simchon Faigler$^{12}$ \\ \\
$^{1}$Center for Astrophysics $|$ Harvard \& Smithsonian, 60 Garden Street, Cambridge, MA 02138, USA\\
$^{2}$Harvard Society of Fellows, 78 Mount Auburn Street, Cambridge, MA 02138\\
$^{3}$Max-Planck Institute for Astronomy, K\"onigstuhl 17, D-69117 Heidelberg, Germany\\
$^{4}$Department of Astronomy, California Institute of Technology, Pasadena, CA 91125, USA\\
$^{5}$Department of Astrophysical Sciences, Princeton University, Princeton, NJ 08544, USA\\
$^{6}$Department of Astronomy, The University of Texas at Austin, 2515 Speedway Boulevard, Austin, TX 78712, USA \\
$^{7}$Center for Computational Astrophysics, Flatiron Institute, 162 Fifth Ave, New York, NY, 10010, USA \\
$^{8}$ Max-Planck-Institut für extraterrestrische Physik, Gießenbachstraße 1, 85748 Garching, Germany\\
$^{9}$Department of Particle Physics and Astrophysics, Weizmann Institute of Science, Rehovot 7610001, Israel \\
$^{10}$MIT-Kavli Institute for Astrophysics and Space Research 77 Massachusetts Ave. Cambridge, MA 02139, USA\\
$^{11}$DTU Space, National Space Institute, Technical University of Denmark, Elektrovej 328, DK-2800 Kgs. Lyngby, Denmark\\
$^{12}$School of Physics and Astronomy, Tel Aviv University, Tel Aviv, 6997801, Israel  \\}

\date{\vspace{-1.0cm}}

\pubyear{2023}

\begin{document}
\label{firstpage}
\pagerange{\pageref{firstpage}--\pageref{lastpage}}
\maketitle

\begin{abstract}
We report spectroscopic and photometric follow-up of a dormant black hole (BH) candidate from {\it Gaia} DR3. The system, which we call Gaia BH2, contains a $\sim$1\,$M_{\odot}$ red giant and a dark companion with mass $M_2 = 8.9\pm 0.3\,M_{\odot}$ that is very likely a BH. The orbital period, $P_{\rm orb} = 1277$ days, is much longer than that of any previously studied BH binary. Our radial velocity (RV) follow-up over a 7-month period spans $>$90\% of the orbit's RV range and is in excellent agreement with the {\it Gaia} solution. UV imaging and high-resolution optical spectra rule out plausible luminous companions that could explain the orbit. The star is a bright ($G=12.3$), slightly metal-poor ($\rm [Fe/H]=-0.22$) low-luminosity giant ($T_{\rm eff}=4600\,\rm K$; $R = 7.8\,R_{\odot}$; $\log\left[g/\left({\rm cm\,s^{-2}}\right)\right] = 2.6$). The binary's orbit is moderately eccentric ($e=0.52$). The giant is enhanced in $\alpha-$elements, with $\rm [\alpha/Fe] = +0.26$, but the system's Galactocentric orbit is typical of the thin disk. We obtained X-ray and radio nondetections of the source near periastron, which support BH accretion models in which the net accretion rate at the horizon is much lower than the Bondi-Hoyle-Lyttleton rate. At a distance of 1.16 kpc, Gaia BH2 is the second-nearest known BH, after Gaia BH1. Its orbit -- like that of Gaia BH1 -- seems too wide to have formed through common envelope evolution. Gaia BH1 and BH2 have orbital periods at opposite edges of the {\it Gaia} DR3 sensitivity curve, perhaps hinting at a bimodal intrinsic period distribution for wide BH binaries. Dormant BH binaries like Gaia BH1 and Gaia BH2 significantly outnumber their close, X-ray bright cousins, but their formation pathways remain uncertain. 
\end{abstract}

\begin{keywords}
binaries: spectroscopic -- stars: black holes 
\vspace{-0.5cm}
\end{keywords}



\section{Introduction}
\label{sec:intro}

The Milky Way is very likely teeming with stellar-mass black holes (BHs). We can see today that it contains a few $\times\, 10^4$ O stars with masses $M\gtrsim 20\,M_{\odot}$ \citep{Garmany1982, Reed2003}, a majority of which will likely leave behind BHs when they die. The lifetime of these massive stars is of order 10 Myr, or 0.1\% the age of the Milky Way. Notwithstanding variations in the Galactic star formation rate with cosmic time, this implies that the Milky Way should contain $\sim 1000$ BHs for every luminous BH progenitor alive today; i.e., a few $\times\,10^7$ BHs. 

Only a tiny fraction of this BH population has been observed. Observations of X-ray binaries have dynamically confirmed the presence of a BH in $\sim$20 systems \citep{Remillard2006}, and another $\sim$50 X-ray transients are suspected to contain BHs \citep{Corral-Santana2016}. In all these systems, there is ongoing mass transfer from a luminous star onto a BH, giving rise to an accretion disk that shines brightly in X-rays. BH X-ray binaries are an intrinsically rare outcome of binary evolution \citep[e.g.][]{PortegiesZwart1997}, and empirical population models suggest that only about 1000 exist in the Milky Way \citep{Corral-Santana2016}. They nevertheless make up the bulk of the {\it observed} BH population, because they are easier to find than BHs that are not accreting. 

The {\it Gaia} mission has opened a new window on the Milky Way's binary star population. By precisely measuring the positions of $\sim$2 billion sources over time, {\it Gaia} can detect subtle wobbles caused by the gravitational effects of binary companions -- even companions that do not emit any light themselves. As of the mission's 3rd data release \citep[``DR3'';][]{GaiaCollaboration2022b}, nearly 500 million sources have been observed in more than 20 well-separated epochs. These data are ultimately expected to yield astrometric binary solutions for of order $10^7$ binaries \citep[e.g.][]{Soderhjelm2004}, but the quality cuts applied to solutions included in DR3 were stringent, so that only $1.7\times 10^5$ full astrometric orbital solutions have been published so far. The DR3 binary catalog nevertheless represents more than an order of magnitude increase in sample size over previous samples of binary orbits and is therefore a promising dataset to search for rare objects. 

{\it Gaia} DR3 has thus far resulted in the identification of one unambiguous BH binary \citep{El-Badry2023, Chakrabarti2022}. That system, Gaia BH1, consists of a Sun-like star in a wide ($P_{\rm orb} =  186$\,day; $a=1.4$\,au) orbit with a $\sim10\,M_{\odot}$ BH companion. The binary's wide orbit is not easily explained by standard binary evolution models. At a distance of only $d=480$ pc, Gaia BH1 is a factor of $\sim$3 closer than the next-nearest known BH, and so the Copernican principle suggests that widely separated BH binaries similar to Gaia BH1 are likely common compared to BHs in X-ray binaries. Most theoretical models also predict that wide binaries with non-accreting BHs should significantly outnumber their accreting cousins \citep[e.g.][]{Breivik2017, Langer2020}. 

{\it Gaia} data are well suited to identify such systems. While Gaia BH1 was the most obvious BH binary candidate published in DR3, \citet{El-Badry2023} also identified a second promising candidate, whose longer orbital period demanded longer-term spectroscopic follow-up than was necessary for Gaia BH1. This paper is focused on that second object, which we refer to as Gaia BH2. Similar to Gaia BH1, the system contains a solar-mass star in a wide orbit around an object we suspect to be a BH. Unlike in Gaia BH1, the luminous star has left the main sequence and is ascending the giant branch. At a distance of 1.16 kpc, the object is likely the second-nearest known BH, after Gaia BH1. 

The remainder of this paper is organized as follows. Section~\ref{sec:disc} describes the system's discovery and {\it Gaia} orbital solution. Most of the observational data and modeling are described in Section~\ref{sec:properties}, including light curves (Section~\ref{sec:lightcurves}), extinction (Section~\ref{sec:extinction}), UV photometry (Section~\ref{sec:swift}), SED modeling (Section~\ref{sec:SEDmod}), mass inference of the giant (Section~\ref{sec:mist}), spectroscopy (Section~\ref{sec:spectroscopy}), orbit modeling (Section~\ref{sec:orbit_modeling}), spectral analysis (Section~\ref{sec:abundances}), Galactic orbit (Section~\ref{sec:galpy}), X-ray observations (Section~\ref{sec:chandra}), and radio observations (Section~\ref{sec:meerkat}). We discuss the nature of the unseen companion in Section~\ref{sec:nature_of_companion} and its possible formation history in Section~\ref{sec:formation_history}. Prospects for detecting and characterizing additional BHs in binaries with {\it Gaia} are discussed in Section~\ref{sec:future}, and we conclude in Section~\ref{sec:concl}. The appendices provide additional details about several aspects of the data and modeling. 

\section{Discovery}
\label{sec:disc}

Being a bright optical source ($G=12.3$), Gaia BH2 appears in many archival photometric catalogs.  To our knowledge, no spectroscopic observations of the source occurred prior to the {\it Gaia} observations.  An orbital solution with a best-fit period of 1352 days -- somewhat longer than the $\sim 1000$ day time baseline for observations included in {\it Gaia} DR3 -- was published in {\it Gaia} DR3 \citep{Arenou2022} with source id 5870569352746779008. We discuss the cadence and phase coverage of the {\it Gaia} observations in Appendix~\ref{sec:appendix_gost}, and the astrometric gooodness of fit in Appendix~\ref{sec:gaia_appendix}. Both are unproblematic. 

Unlike Gaia BH1, which has a purely astrometric solution (\texttt{nss\_solution\_type = Orbital}), Gaia BH2 has a joint astrometric + single-lined spectroscopic solution (\texttt{nss\_solution\_type = AstroSpectroSB1}). The reason for this difference is that individual-epoch RVs are only derived for sources brighter than $G_{\rm RVS}=12$, where $G_{\rm RVS}$ is a source's magnitude in the {\it Gaia} RVS bandpass. Gaia BH2 has $G_{\rm RVS}=11.11$, and RVs were measured (but not published) in 34 unique transits. Gaia BH1 has $G_{\rm RVS}=12.83$, and thus no individual-epoch RVs.

We identified Gaia BH2 as a promising BH candidate in the course of a broader program to spectroscopically follow-up {\it Gaia} DR3 binaries suspected to contain compact objects. The astrometric selection that first brought the system to our attention is described in Appendix E of \citet{El-Badry2023}. The {\it Gaia} solution has a period of $P_{\rm orb} = 1352\pm 46$ days. The astrometric mass function implied by the astrometric data alone is $f\left(M_{2}\right)_{{\rm ast}}=\left(a_{0}/\varpi\right)^{3}\left(P_{{\rm orb}}/{\rm yr}\right)^{-2}=6.72\pm0.5\,M_{\odot}$; here $a_0$ and $\varpi$ represent the semi-major axis of the photocenter and the parallax. Assuming a mass of $1\,M_{\odot}$ for the luminous star (Section~\ref{sec:mist}), this implied an unseen companion mass of $M_2 \approx 8.4\pm 0.5\,M_{\odot}$. Since this far exceeds the apparent photometric mass of the luminous primary, we initiated a spectroscopic follow-up campaign,  obtaining our first spectrum of the source in August 2022. 
 
In \citet{El-Badry2023}, we reported that early follow-up data were consistent with the {\it Gaia} solution. However, since the data in-hand then covered only a small fraction of the orbit's predicted RV range, we cautioned that continued RV monitoring was required to better assess the reliability of the {\it Gaia} solution. We have now carried out the necessary observations.

{\it Gaia} data for the source were also analyzed by \citet{Tanikawa2022}, who were the first to draw attention to the source as a BH candidate in a September 2022 preprint. They concluded that the {\it Gaia} orbital solution makes the source the most promising BH candidate among the $\sim$64000 binary solutions in {\it Gaia} DR3 for which both time-resolved spectroscopic and astrometric data are available. Their work showed convincingly that the {\it Gaia} data suggest the presence of an unseen massive companion, which would be most simply explained by a BH. These authors did not obtain any follow-up data, so their conclusions depend entirely on the reliability of the {\it Gaia} solution. 

\section{Properties of the source}
\label{sec:properties}
\begin{figure*}
    \centering
    \includegraphics[width=\textwidth]{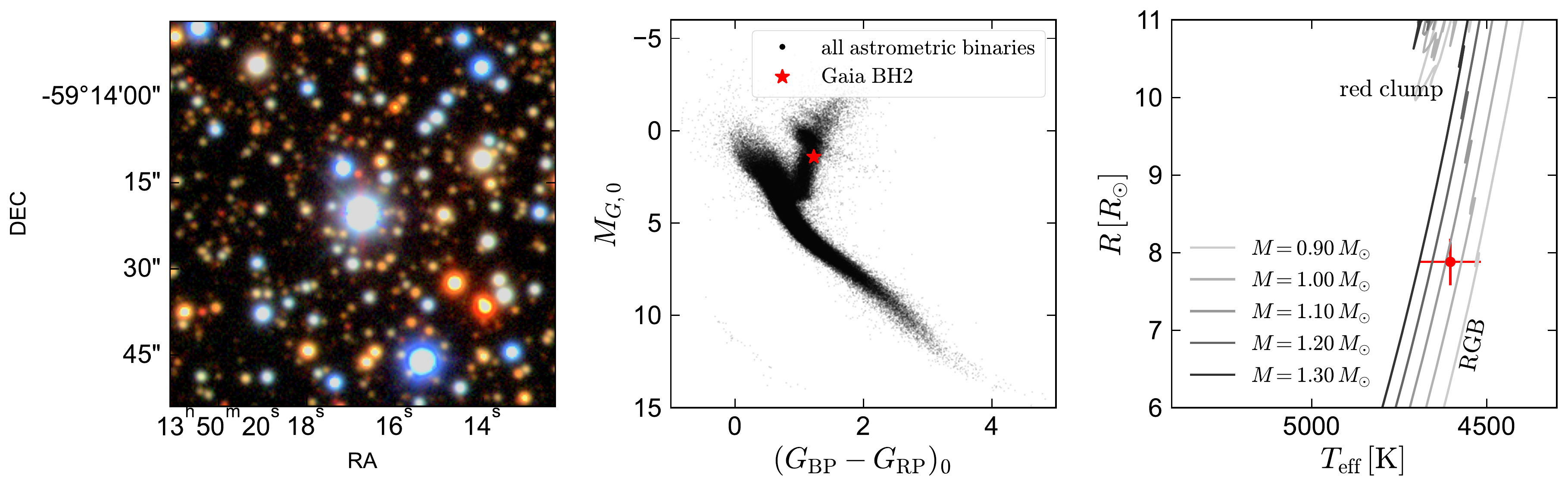}
    \caption{Properties of the luminous star. Left panel shows a 1.1-arcmin  $grz$-band postage stamp from the DECaPS survey. Gaia BH2 is the bright (saturated) source at the center. Middle panel shows the source's position on the dereddened {\it Gaia} color-magnitude diagram, with the full {\it Gaia} DR3 astrometric binary sample shown for context. Right panel compares the star's inferred temperature and radius to single-star MIST evolutionary models. The optical source appears to be an evolved $\sim 1\,M_{\odot}$ star on the lower giant branch. }
    \label{fig:mosaic}
\end{figure*}

Gaia BH2 is a source in Centaurus ($\alpha=$\,13:50:16.7; $\delta=-$59:14:20), near the Galactic plane ($l = 310.4, b=2.8$), with $G=12.3$. Its basic properties are summarized in Figure~\ref{fig:mosaic}. The field is relatively crowded, with 17 {\it Gaia}-detected neighbors within 10 arcsec, though all are at least 3.5 magnitudes fainter than the source itself. The nearest {\it Gaia}-detected neighbor is a $G=17.95$ source at a distance of 1.77 arcsec. This source is unresolved in the postage stamp shown in Figure~\ref{fig:mosaic}, which is from the DECaPS survey \citep{Schlafly2018, Saydjari2022}. Its astrometry is inconsistent with being bound to Gaia BH2.

The {\it Gaia} astrometric solution places Gaia BH2 at a distance $d = 1.16 \pm 0.02$\,kpc. We do not attempt to correct this distance for the {\it Gaia} parallax zeropoint \citep{Lindegren2021b}, since there is little reason to expect that this zeropoint -- inferred from single-star astrometric solutions -- applies to the binary solutions.
On the color-magnitude diagram, Gaia BH2 appears on the lower giant branch, just below the red clump. Our constraints on the star's temperature and radius (Section~\ref{sec:SEDmod}) imply a mass near $1\,M_{\odot}$, as shown in the right panel of Figure~\ref{fig:mosaic}.

\subsection{Light curves}
\label{sec:lightcurves}
We inspected the ASAS-SN $V-$ and $g-$band light curves of the source \citep{Kochanek_2017}, which contain more than 3000 photometric epochs over a 7-year period, with a typical uncertainty of 0.02 mag. This did not reveal any significant periodic or long-term photometric variability: the optical source is constant at the $\gtrsim 0.01$ mag level. 

Gaia BH2 was observed by {\it TESS} \citep{Ricker2015} during sectors 11 and 38. The source is sufficiently bright ($T\approx 11.5$) that it may be possible to measure asteroseismic parameters from the light curve \citep{Stello2022}. We extracted light curves from the  {\it TESS}  full frame images using \texttt{eleanor} \citep{Feinstein2019} and analyzed their power spectra following the methods described by \citet{Stello2022}; see Appendix~\ref{sec:tess} for details. We found a marginally significant detection of $\nu_{\rm max}=61\,\mu \rm Hz$ in both sectors. If reliable, this value can be translated into a constraint on the giant's surface gravity using asteroiseismic scaling relations \citep[e.g.][]{Chaplin2013}. Together with the spectroscopically-measured effective temperature, it implies a surface gravity $\log\left[g/\left({\rm cm\,s^{-2}}\right)\right]\approx 2.67$, consistent with the spectroscopic $\log g$ (Section~\ref{sec:abundances}). Given the marginal nature of the detection, we did not include it in our subsequent analysis. 

\subsection{Extinction}
\label{sec:extinction}
We estimate the extinction to Gaia BH2 using the 3D dust map from \citet{Lallement2022}, which predicts a modest integrated extinction of $A_{550\,\rm nm}=0.58\,\rm mag$ at $d=1.16$\,kpc, corresponding to $E(B-V)\approx 0.2$ mag. The $G-$band extinction inferred by the {\it Gaia} \texttt{GSP-Phot} pipeline \citep{Andrae2022} is $A_G= 0.70$, corresponding to $E(B-V)\approx 0.26$, but we expect the value from the 3D dust map to be more reliable for stars in this evolutionary phase. We adopt $E(B-V)=0.2\pm 0.03$ mag in our SED modeling. 

\subsection{Ultraviolet observation}
\label{sec:swift}
Gaia BH2 does not have a published UV magnitude and is outside the published GALEX footprint. Measuring a UV magnitude or upper limit for the source is important for constraining possible flux contributions from a luminous secondary, so we obtained a 1625s observation of the source with the UVOT photometer \citep{Roming2005} on board the {\it Neil Gehrels Swift Observatory} in September 2022 (ToO ID 17935). We used the UVM2 band, with an effective wavelength of 2246\,\AA, and also obtained simultaneous X-ray observations using XRT \citep{Burrows2005}.

We analyzed the UVM2 image using the \texttt{uvotsource} routine distributed with \texttt{Heasoft}, using a 4 arcsec aperture centered on the {\it Gaia} source and an 8 arcsec background aperture in a nearby region with no obvious sources in the UVM2 image (ra = 207.558463, dec = -59.231955). Gaia BH2 is detected in the UVM2 image only with 2.7$\sigma$ significance. A low-significance detection turns out to be sufficient for our purposes, since the main goal was to rule out a UV-bright companion. The AB magnitude in the UVM2 band is $22.20\pm 0.41$ mag, corresponding to a flux density of $(2.9 \pm 1.1)\times 10^{-17}\,\rm erg\,s^{-1}\,cm^{-2}$.

The source was not detected in the simultaneous XRT observation. We did not analyze the XRT data further, since our {\it Chandra} observations (Section~\ref{sec:chandra}) provide significantly deeper X-ray constraints.

\begin{figure*}
    \centering
    \includegraphics[width=\textwidth]{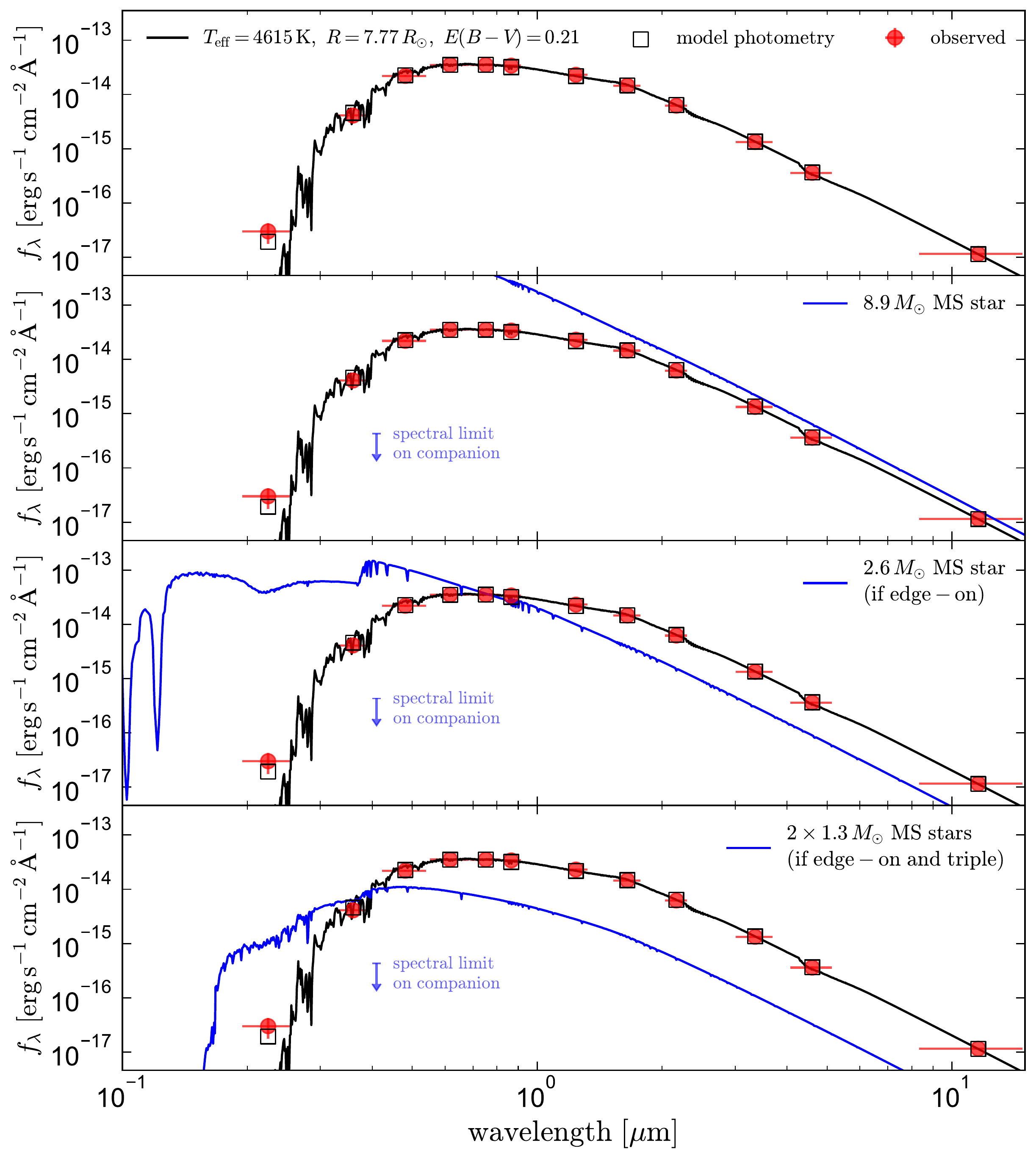}
    \caption{Spectral energy distribution of the Gaia BH2 system. In all panels, red points show observed photometry. Black line and open squares show the continuous SED and model photometry for the best-fit single-star model, whose parameters are listed in the top panel. 2nd panel shows the SED of an $8.9\,M_{\odot}$ main-sequence star; this would be 200 times more luminous than the observed source, and is thus ruled out. In the third and fourth panels, we consider a scenario where the {\it Gaia} astrometric inclination constraint (which implies $M_2\sim 8.9\,M_{\odot}$) is spurious. In this case, the minimum plausible companion mass is $M_2\sim2.6\,M_{\odot}$, corresponding to an edge-on orbit (Figure~\ref{fig:mass_fn}). Main-sequence companions of this mass would still be brighter than the observed source in the UV, both for the case of a single $2.6\,M_{\odot}$ companion (3rd panel) and for a $1.3+1.3\,M_{\odot}$ inner binary (4th panel). These scenarios would also conflict with limits on the companion flux from the optical spectra (Section~\ref{sec:uves_luminous}) which are shown with a blue upper limit. }
    \label{fig:seds}
\end{figure*}

\subsection{SED model}
\label{sec:SEDmod}

We constructed the source's broadband spectral energy distribution (SED) by combining the {\it Swift} UVM2 magnitude, synthetic SDSS $ugriz$  photometry constructed from {\it Gaia} BP/RP spectra \citep{GaiaCollaboration2022}, 2MASS $JHK$ photometry \citep[][]{Skrutskie_2006}, and WISE $W1\,W2\,W3$ photometry \citep[][]{Wright_2010}. We set an uncertainty floor of 0.02 mag in all bands to allow for photometric calibration issues and imperfect models. We then fit the SED with single-star models to infer the temperature and radius of the red giant. 
We predict bandpass-integrated magnitudes using empirically calibrated model spectral from the BaSeL library \citep[v2.2;][]{Lejeune1997, Lejeune1998}. We assume a \citet{Cardelli_1989} extinction law with $R_V =3.1$ and adopt a prior on the reddening $E(B-V) = 0.20\pm 0.03$. We use \texttt{pystellibs}\footnote{\href{https://mfouesneau.github.io/pystellibs/}{https://mfouesneau.github.io/pystellibs/}} to interpolate between model SEDs, and \texttt{pyphot}\footnote{\href{https://mfouesneau.github.io/pyphot/}{https://mfouesneau.github.io/pyphot/}}  to calculate synthetic photometry. We then fit the SED using \texttt{emcee} \citep{emcee2013} to sample from the posterior, with the temperature, radius, metallicity, and reddening sampled as free parameters. 

The results are shown in the top panel of Figure~\ref{fig:seds}. A single-star model yields a reasonably good fit to the data, with $\chi^2/N_{\rm data} = 1.16$, where $N_{\rm data}$ is the number of photometric points. The best-fit SED model has $T_{\rm eff}=4615\,\rm K$ and $R = 7.77\,R_{\odot}$, with a bolometric luminosity of $L\approx 25\,L_{\odot}$. Modeling of the source's high-resolution spectrum (Section~\ref{sec:abundances}) leads to a very similar constraint on $T_{\rm eff}$. The high-precision of the {\it Gaia} parallax ($\varpi/\sigma_{\varpi}=47$) allows us to constrain the star's radius within $\approx 3\%$ and the luminosity within $\approx 6\%$. 

The lack of significant UV excess leads to stringent limits on any possible main-sequence companions: if the companion were a normal star, it would be impossible to obtain a good fit to the SED with a single-star model \citep[e.g.][]{El-Badry2022_gaia_algols}. Luminous stars that are massive enough to explain the observed astrometry and RVs are predicted to outshine the giant in the UV and contribute significantly to the optical spectrum. These scenarios are explored in the lower panels of Figure~\ref{fig:seds} and discussed further in Section~\ref{sec:nature_of_companion}.

\subsection{Luminous star mass and evolutionary state}
\label{sec:mist}
We compare the measured radius and effective temperature of the luminous star to MIST single-star evolutionary models \citep{Choi2016}. We use solar-scaled models with $Z=0.012$ to account approximately for the star's measured sub-solar iron abundance and enhancement of $\alpha-$elements (see \citealt{Fu2018}; Gaia BH2 has $\rm [Fe/H]=-0.22$ and $[\rm \alpha/Fe] = +0.26$, as described in Section~\ref{sec:abundances}). 
These models are shown in the right panel of Figure~\ref{fig:mosaic}.

Fitting the measured temperature and radius implies a luminous star mass of $1.07\pm 0.19\,M_{\odot}$. The uncertainty is dominated by uncertainty in the observed effective temperature, though uncertainties in the stellar models likely contribute at a similar level \citep[e.g.][]{Joyce2022}. The corresponding age range is about 5-13 Gyr. In all plausible models, the star is on the lower red giant branch and has recently begun to expand following core hydrogen exhaustion; it is {\it not} a core helium burning red clump star that has already reached the tip of the giant branch and experienced a helium flash.

\subsubsection{Could the star be a low-mass stripped giant?}
We considered whether the giant might be a product of recent envelope stripping, in which case it might have significantly lower mass than implied by single-star evolutionary models \citep[e.g.][]{El-Badry2022_unicorns, El-Badry2022_gaia_algols}. We consider such a scenario unlikely given the binary's wide and non-circular orbit, and the lack of evidence for ongoing mass transfer. In the current orbit, the star would overflow its Roche lobe at periastron only if it had a radius $R\gtrsim 100\,R_{\odot}$. This essentially precludes a low-mass ($M\lesssim 0.5\,M_{\odot}$) stripped giant, because a giant of that mass would not have reached a sufficiently large radius to overflow its Roche lobe in the Gaia BH2 system \citep[e.g.][]{Rappaport1995}. A past period of mass transfer from the giant to the companion would also be expected to have circularized the orbit. Our analysis of the star's spectrum reveals no excess absorption or emission in H$\alpha$ (Section~\ref{sec:abundances}), as has been commonly found in giants with ongoing or recently-terminated mass transfer \citep{Jayasinghe2021, Jayasinghe2022}. Finally, the spectroscopic $\log g$ (Section~\ref{sec:abundances}) and the radius constraint from the SED fit imply a giant mass of $M_\star = 1.17^{+0.88}_{-0.50}\,M_{\odot}$; i.e., a 1$\sigma$ lower limit of $0.67\,M_{\odot}$.

\subsection{Spectroscopy}
\label{sec:spectroscopy}
To measure atmospheric parameters, abundances, and radial velocities, we obtained multi-epoch spectra of Gaia BH2 using two instruments. 

\subsubsection{FEROS}
We observed Gaia BH2 43 times with the Fiberfed Extended Range Optical Spectrograph \citep[FEROS;][]{Kaufer1999} on the 2.2m ESO/MPG telescope at La Silla Observatory (programs P109.A-9001 and P110.9014). The first several observations used $2\times 2$ binning to reduce readout noise at the expense of spectral resolution; the remainder used $1\times 1$ binning. The resulting spectra have resolution $R\approx 40,000$ ($2\times 2$ binning) and $R\approx 50,000$ ($1\times 1$ binning). Most of our observations used 1200s or 900s exposures. The typical SNR at 5800\,\AA\, is 15-30.

We reduced the data using the CERES pipeline \citep{Brahm2017}, which performs bias-subtraction, flat fielding, wavelength calibration, and optimal extraction. The pipeline measures and corrects for small shifts in the wavelength solution during the course a night via simultaneous observations of a ThAr lamp with a second fiber. Several minor modifications to the pipeline were necessary to obtain good performance with $2\times 2$ binning. We first calculate RVs by cross-correlating a template spectrum with each order individually and then report the mean RV across 15 orders with wavelengths between 4500 and 6700\,\AA. We calculate  the uncertainty on this mean RV from the dispersion between orders; i.e., $\sigma_{{\rm RV}}\approx{\rm std}\left({\rm RVs}\right)/\sqrt{15}$.

\subsubsection{UVES}
We observed Gaia BH2 5 times with the Ultraviolet-Visible Echelle Spectrograph \citep[UVES;][]{Dekker2000} mounted on the 8.2m UT2 telescope at the VLT on Cerro Paranal (program 2110.D-5024). We used Dichroic 2 with central wavelengths of 437 and 760 nm on the blue and red arms, providing spectral coverage of the wavelength ranges from 3730-4990 and 5670-9450\,\AA. We took 600s exposures and used 0.4 and 0.3 arcsec slits on the blue and red arms, respectively, which yielded resolutions of 80,000 and 110,000 with SNR $\sim$50 at 6500\,\AA. 

We reduced the data using the ESO Reflex pipeline \citep{Freudling2013} with standard calibrations. This performs bias-subtraction, flat fielding, wavelength calibration using afternoon ThAr arcs, optimal extraction, and order merging. We verified the stability of the wavelength solution using telluric absorption lines, measured RVs via cross-correlation of the merged red-arm spectra with a template, and adopted a per-epoch RV uncertainty of 0.1 $\rm km\,s^{-1}$ to account for drifts in the wavelength solution between science exposures and afternoon arcs. Near-simultaneous FEROS and UVES observations allowed us to verify good agreement between the two RV scales at the $\approx 0.1\,\rm km\,s^{-1}$ level.

\subsubsection{Radial velocities}
\label{sec:rvs}

\begin{table}
\begin{tabular}{llll}
HJD UTC & RV ($\rm km\,s^{-1}$) & Instrument & SNR \\
\hline
2459814.4989 & $10.99\pm 0.04$  & FEROS ($2 \times 2$) & 27 \\
2459916.8408 & $-2.05\pm 0.04$  & FEROS ($2 \times 2$) & 26 \\
2459917.8428 & $-2.24\pm 0.04$  & FEROS ($2 \times 2$) & 31 \\
2459918.8523 & $-2.60\pm 0.03$  & FEROS ($2 \times 2$) & 35 \\
2459919.8432 & $-2.84\pm 0.04$  & FEROS ($2 \times 2$) & 28 \\
 2459920.8374 & $-3.06\pm 0.08$  & FEROS ($2 \times 2$)& 15 \\
2459921.8421 & $-3.40\pm 0.03$  & FEROS ($2 \times 2$) & 33 \\
2459923.8343 & $-4.08\pm 0.09$  & FEROS ($2 \times 2$) & 11 \\
2459924.8366 & $-4.21\pm 0.03$  & FEROS ($1 \times 1$) & 19 \\
2459932.8561 & $-6.69 \pm 0.03$  & FEROS ($1 \times 1$) & 19 \\
2459935.8419 & $-7.68\pm 0.1$  & UVES & 50 \\
2459940.8282 & $-9.43\pm 0.03$  & FEROS ($1 \times 1$) & 20 \\
2459947.8362 & $-12.02\pm 0.04$  & FEROS ($1 \times 1$)  & 25 \\
2459957.8404 & $-15.95\pm 0.1$  & UVES & 50 \\
2459957.8637 & $-15.89\pm 0.05$  & FEROS ($1 \times 1$) & 18 \\
2459958.8580 & $-16.29\pm 0.05$  & FEROS ($1 \times 1$) & 21 \\
2459959.8584 & $-16.71\pm 0.03$  & FEROS ($1 \times 1$) & 19 \\
2459960.8611 & $-17.16\pm 0.04$  & FEROS ($1 \times 1$) & 15 \\
2459961.8795 & $-17.52\pm 0.04$  & FEROS ($1 \times 1$) & 15 \\
2459962.8667 & $-17.95\pm 0.04$  & FEROS ($1 \times 1$) & 16 \\
2459965.8774 & $-19.26\pm 0.03$  & FEROS ($1 \times 1$) & 23 \\
2459966.8686 & $-19.58\pm 0.03$  & FEROS ($1 \times 1$) & 23 \\
2459967.8619 & $-19.98\pm 0.03$  & FEROS ($1 \times 1$) & 21 \\
2459972.7330 & $-22.10\pm 0.10$  & UVES  & 45 \\
2459984.8437 & $-26.82\pm 0.03$  & FEROS ($1 \times 1$)  & 23 \\
2459985.8316 & $-27.18\pm 0.03$  & FEROS ($1 \times 1$)  & 20 \\
2459986.8430 & $-27.56\pm 0.04$  & FEROS ($1 \times 1$)  & 16 \\
2459987.8336 & $-27.93\pm 0.03$  & FEROS ($1 \times 1$)  & 20 \\
2459989.7996 & $-28.60\pm 0.03$  & FEROS ($1 \times 1$)  & 21 \\
2459990.7861 & $-28.96\pm 0.03$  & FEROS ($1 \times 1$)  & 22 \\
2459992.8014 & $-29.66\pm 0.03$  & FEROS ($1 \times 1$)  & 21 \\
2459994.7108 & $-30.23\pm 0.1$   & UVES & 40 \\
2459997.8192 & $-31.28\pm 0.04$   & FEROS ($1 \times 1$) & 16 \\
2459999.8278 & $-31.88\pm 0.03$   & FEROS ($1 \times 1$) & 20 \\
2460000.8586 & $-32.15\pm 0.04$   & FEROS ($1 \times 1$) & 18 \\
2460001.8520 & $-32.50\pm 0.04$   & FEROS ($1 \times 1$) & 21 \\
2460002.8348 & $-32.76\pm 0.04$   & FEROS ($1 \times 1$) & 21 \\
2460003.8399 & $-33.05\pm 0.03$   & FEROS ($1 \times 1$) & 21 \\
2460004.8448 & $-33.31\pm 0.03$   & FEROS ($1 \times 1$) & 22 \\
2460005.8399 & $-33.55\pm 0.03$   & FEROS ($1 \times 1$) & 21 \\
2460006.8482 & $-33.82\pm 0.03$   & FEROS ($1 \times 1$) & 21 \\
2460012.7997 & $-35.14\pm 0.03$   & FEROS ($1 \times 1$) & 21 \\
2460013.8823 & $-35.38\pm 0.03$   & FEROS ($1 \times 1$) & 20 \\
2460014.8624 & $-35.55\pm 0.03$   & FEROS ($1 \times 1$) & 19 \\
2460015.8741 & $-35.71\pm 0.03$   & FEROS ($1 \times 1$) & 19 \\
2460017.8436 & $-36.06\pm 0.03$   & FEROS ($1 \times 1$) & 21 \\
2460018.6749 & $-36.22\pm 0.1$   & UVES & 45 \\
2460018.8868 & $-36.27\pm 0.04$   & FEROS ($1 \times 1$) & 20 \\

\hline

\end{tabular}
\caption{Radial velocities.}
\label{tab:rvs}
\end{table}

Our follow-up RVs are shown in Figure~\ref{fig:rvfig} and listed in Table~\ref{tab:rvs}. In total, we obtained 48 RVs between August 2022 and March 2023. All our observations have a precision better than 0.1 $\rm km\,s^{-1}$. The precision of the FEROS RVs is somewhat higher than that of the UVES data because of the precise wavelength calibration enabled by simultaneous arcs. 

\begin{table}
\centering
\caption{Physical parameters and 1$\sigma$ uncertainties for both components of Gaia BH2. We compare constraints on the orbit based on both {\it Gaia} and our RVs (3rd block), {\it Gaia} alone (4th block), and our RVs alone (5th) block. The ``our RVs alone'' constraints adopt a prior of $P_{\rm orb} < 2000$ days. Flat and broad priors are used for all other constraints.}
\begin{tabular}{lll}
\hline\hline
\multicolumn{3}{l}{\bf{Properties of the unresolved source}}   \\ 
Right ascension & $\alpha$\,[deg] & 207.56971624 \\
Declination & $\delta$\,[deg] & -59.2390050 \\
Apparent magnitude & $G$\,[mag] & 12.28 \\
Parallax & $\varpi$\,[mas] & $ 0.859 \pm 0.018 $ \\
Proper motion in RA & $\mu_{\alpha}^{*}$\,[$\rm mas\,yr^{-1}$] & $ -10.48 \pm 0.10 $ \\
Proper motion in Dec & $\mu_{\delta}$\,[$\rm mas\,yr^{-1}$] & $ -4.61 \pm 0.06 $ \\
Tangential velocity & $v_{\perp}\,\left[{\rm km\,s^{-1}}\right]$ & $ 64.4 \pm 0.7$ \\ 
Extinction & $E(B-V)$\,[mag] & $ 0.20 \pm 0.03 $ \\

\hline
\multicolumn{3}{l}{\bf{Parameters of the red giant}}  \\ 
Effective temperature & $T_{\rm eff}$\,[K] & $4604 \pm 87 $ \\
Surface gravity   & $\log(g/(\rm cm\,s^{-2}))$  & $2.71 \pm 0.24$  \\
Projected rotation velocity & $v\sin i$\,[km\,s$^{-1}$] &  $< 1.5$ \\
Radius & $R_\star\,[R_{\odot}]$ & $7.77 \pm 0.25$  \\ 
Bolometric luminosity & $L_\star\,[L_{\odot}]$ & $24.6\pm 1.6$ \\ 
Mass &  $M_\star\,[M_{\odot}]$ & $1.07 \pm 0.19$ \\
Metallicity &  $\rm [Fe/H]$ & $-0.22 \pm 0.02$ \\
$\alpha-$abundance &  $\rm [\alpha/Fe]$ & $0.26 \pm 0.05$ \\
Abundance pattern &  $\rm [X/Fe]$ & Table~\ref{tab:bacchus} \\

\hline
\multicolumn{3}{l}{\bf{Parameters of the orbit ({\it Gaia} + our RVs)}}  \\ 
Orbital period & $P_{\rm orb}$\,[days] & $ 1276.7 \pm 0.6 $ \\
Semi-major axis & $a$\,[au] & $ 4.96 \pm 0.08 $ \\
Photocenter semi-major axis & $a_0$\,[mas] & $ 3.719 \pm 0.014 $ \\
Eccentricity & $e$ & $ 0.5176 \pm 0.0009 $ \\
Inclination  & $i$\,[deg] & $ 34.87 \pm 0.34 $ \\
Periastron time & $T_p$\,[JD-2457389] & $49.3 \pm 1.4 $ \\
Ascending node angle & $\Omega$\,[deg] & $266.9 \pm 0.5 $ \\
Argument of periastron & $\omega$\,[deg] & $130.9 \pm 0.4 $ \\
Black hole mass & $M_2$\,[$M_{\odot}$] & $8.94 \pm 0.34$ \\
Center-of-mass RV & $\gamma$\,[$\rm km\,s^{-1}$] & $-4.22 \pm 0.13$ \\
RV semi-amplitude & $K_\star$\,[$\rm km\,s^{-1}$] & $25.23 \pm 0.04$ \\
RV mass function & $f\left(M_{2}\right)_{{\rm RVs}}$\,[$\rm M_{\odot}$] & $1.331 \pm 0.008$ \\

\hline
\multicolumn{3}{l}{\bf{Parameters of the orbit ({\it Gaia} only)}}  \\ 
Orbital period & $P_{\rm orb}$\,[days] & $ 1300 \pm 26 $ \\
Semi-major axis & $a$\,[au] & $ 5.05 \pm 0.12 $ \\
Photocenter semi-major axis & $a_0$\,[mas] & $ 3.79 \pm 0.06 $ \\
Eccentricity & $e$ & $0.515 \pm 0.01 $ \\
Inclination  & $i$\,[deg] & $ 35.7 \pm 0.7 $ \\
Periastron time & $T_p$\,[JD-2457389] & $49.4 \pm 3.1 $ \\
Ascending node angle & $\Omega$\,[deg] & $266.4 \pm 1.0 $ \\
Argument of periastron & $\omega$\,[deg] & $131.2 \pm 1.6 $ \\
Black hole mass & $M_2$\,[$M_{\odot}$] & $9.1 \pm 0.7 $ \\
Center-of-mass RV & $\gamma$\,[$\rm km\,s^{-1}$] & $-3.7 \pm 0.6 $ \\
RV semi-amplitude & $K_\star$\,[$\rm km\,s^{-1}$] & $25.8 \pm 0.8$ \\
RV mass function & $f\left(M_{2}\right)_{{\rm RVs}}$\,[$\rm M_{\odot}$] & $1.45 \pm 0.145$ \\
\hline
\multicolumn{3}{l}{\bf{Parameters of the orbit (our RVs only)}} \\
Orbital period & $P_{\rm orb}$\,[days] & $1476^{+321}_{-297} $ \\
Eccentricity & $e$ & $0.56^{+0.05}_{-0.06} $ \\
RV semi-amplitude & $K_\star$\,[$\rm km\,s^{-1}$] & $25.38^{+0.18}_{-0.24}$ \\
RV mass function & $f\left(M_{2}\right)_{{\rm RVs}}$\,[$\rm M_{\odot}$] & $1.44^{+0.14}_{-0.16} $ \\
\hline
\end{tabular}
\begin{flushleft}

\label{tab:system}
\end{flushleft}
\end{table}

\subsection{Orbit modeling}
\label{sec:orbit_modeling}

We jointly fit our follow-up RVs and the astrometry+RV constraints from {\it Gaia} with a model that has 14 free parameters: the five standard astrometric parameters ($\alpha, \delta,\mu_{\alpha}^*,\mu_{\delta},$ and $\varpi$); and the binary parameters period, eccentricity, inclination, angle of the ascending node $\Omega$, argument of periastron $\omega$, periastron time, center-of-mass velocity, luminous star mass $M_{\star}$, and companion mass $M_2$. For each call to the likelihood function, we predict the semi-major axis in physical units, $a=\left[P_{\rm orb}^{2}G\left(M_{\star}+M_{2}\right)/\left(4\pi^{2}\right)\right]^{1/3}$. We then calculate the angular semi-major axis of the photocenter ($a_0$), and that of the primary ($a_1$; i.e., the star whose RVs are being measured):
\begin{align}
    a_{1}	&=\frac{a}{d}\left(\frac{q}{1+q}\right) \\
a_{0}	&=\frac{a}{d}\left(\frac{q}{1+q}-\frac{\epsilon}{1+\epsilon}\right).
\end{align}
Here, $q=M_2/M_\star$ is the mass ratio and $\epsilon$ is the $G-$band flux ratio of the companion to the primary. $d$ is the distance, which we compute as $d\left[{\rm kpc}\right]=1/\varpi\left[{\rm mas}\right]$. For a dark companion ($\epsilon=0$), $a_1=a_0$. We then predict the six Thiele-Innes parameters:
\begin{align}
\label{eq:Apred}
    A	&=a_{0}\left(\cos\omega\cos\Omega-\sin\omega\sin\Omega\cos i\right) \\
B	&=a_{0}\left(\cos\omega\sin\Omega+\sin\omega\cos\Omega\cos i\right) \\
F	&=-a_{0}\left(\sin\omega\cos\Omega+\cos\omega\sin\Omega\cos i\right) \\
G	&=-a_{0}\left(\sin\omega\sin\Omega-\cos\omega\cos\Omega\cos i\right) \\
C	&=\frac{a_{1}}{\varpi}\sin\omega\sin i \\
\label{eq:Hpred}
H	&=\frac{a_{1}}{\varpi}\cos\omega\sin i.
\end{align}
Note that in the convention used here and in the {\it Gaia} archive, the astrometric parameters $A,B,F$, and $G$ have angular units (mas), while the spectroscopic parameters $C$ and $H$ have physical units (au). In the {\it Gaia} data processing, $C$ and $H$ are constrained by the measured RVs, while $A,B,F$, and $G$ are constrained by the astrometry \citep[e.g.][]{Pourbaix2022}. 


For each call to the likelihood function, we construct the predicted vector of {\it Gaia}-constrained parameters:

\begin{equation}
    \label{eq:gaia_vec}
    \boldsymbol \theta_{\rm Gaia}=\left[\alpha,\delta,\varpi,\mu_{\alpha}^{*},\mu_{\delta},A,B,F,G,C,H,\gamma,e,P_{{\rm orb}},T_{p}\right].
\end{equation}
We then calculate a likelihood term that quantifies the difference between these quantities and the {\it Gaia} constraints: 
\begin{equation}
    \label{eq:lnL_ast}
    \ln L_{{\rm Gaia}}=-\frac{1}{2}\left(\boldsymbol  \theta_{{\rm Gaia}}-\boldsymbol \mu_{{\rm Gaia}}\right)^{\intercal}\boldsymbol\Sigma_{{\rm Gaia}}^{-1}\left(\boldsymbol \theta_{{\rm Gaia}}-\boldsymbol \mu_{{\rm Gaia}}\right).
\end{equation}
Here $\boldsymbol \mu_{{\rm Gaia}}$ and $\boldsymbol \Sigma_{\rm Gaia}$ represent the vector of best-fit parameters constrained by {\it Gaia} and their covariance matrix, which we construct from the \texttt{corr\_vec} parameter reported in the {\it Gaia} archive. 

We additionally predict the RVs of the luminous star at the array of times $t_i$ at which we obtained spectra. This leads to a radial velocity term in the likelihood, 

\begin{equation}
    \label{eq:lnL_RVs}
    \ln L_{{\rm RVs}}=-\frac{1}{2}\sum_{i}\frac{\left({\rm RV_{{\rm pred}}}\left(t_{i}\right)-{\rm RV}_{i}\right)^{2}}{\sigma_{{\rm RV,}i}^{2}},
\end{equation}
where ${\rm RV}_i$ and ${\rm RV}_{{\rm pred}}\left(t_{i}\right)$ are the measured and predicted RVs, with their uncertainties $\sigma_{\rm RV,i}$. The predicted RVs are calculated by solving Kepler's equation using standard methods.
The full likelihood is then given by
\begin{equation}
    \label{eq:lnL_tot}
    \ln L = \ln L_{{\rm Gaia}} +  \ln L_{{\rm RVs}}.
\end{equation}

We use flat priors on all parameters except $M_{\star}$, for which we use a normal distribution informed by isochrones, $M_{\star}/M_{\odot}\sim \mathcal{N}(1.07,0.19)$. We sample from the posterior using \texttt{emcee} \citep[][]{emcee2013} with 64 walkers, taking 3000 steps per walker after a burn-in period of 3000 steps. The results of the fit are summarized in Table~\ref{tab:system} (3rd block).

We initially considered including the flux ratio $\epsilon$ as a free parameter. But in this case we found the posterior for $\epsilon$ to pile up against 0, with a 1$\sigma$ upper limit of $\epsilon < 0.11$. That is, a luminous companion is disfavored by the astrometry because it would produce too small of a photocenter semi-major axis, independent of any arguments about the expected mass-luminosity relation of plausible luminous secondaries. In subsequent analysis, we therefore fixed $\epsilon=0$ in order to avoid introducing a bias due to unrealistically large flux contributions from the secondary, which we suspect is truly dark.

\subsubsection{Gaia-only constraints}
\label{sec:gaia_only}

In order to explore the relative importance of our follow-up RVs and the {\it Gaia} data for our final constraints, we also repeat the fit without the follow-up RVs; i.e., simply removing the $\ln L_{\rm RVs}$ term from Equation~\ref{eq:lnL_tot}. We list the resulting constraints in Table~\ref{tab:system} (4th block).

\begin{figure*}
    \centering
    \includegraphics[width=\textwidth]{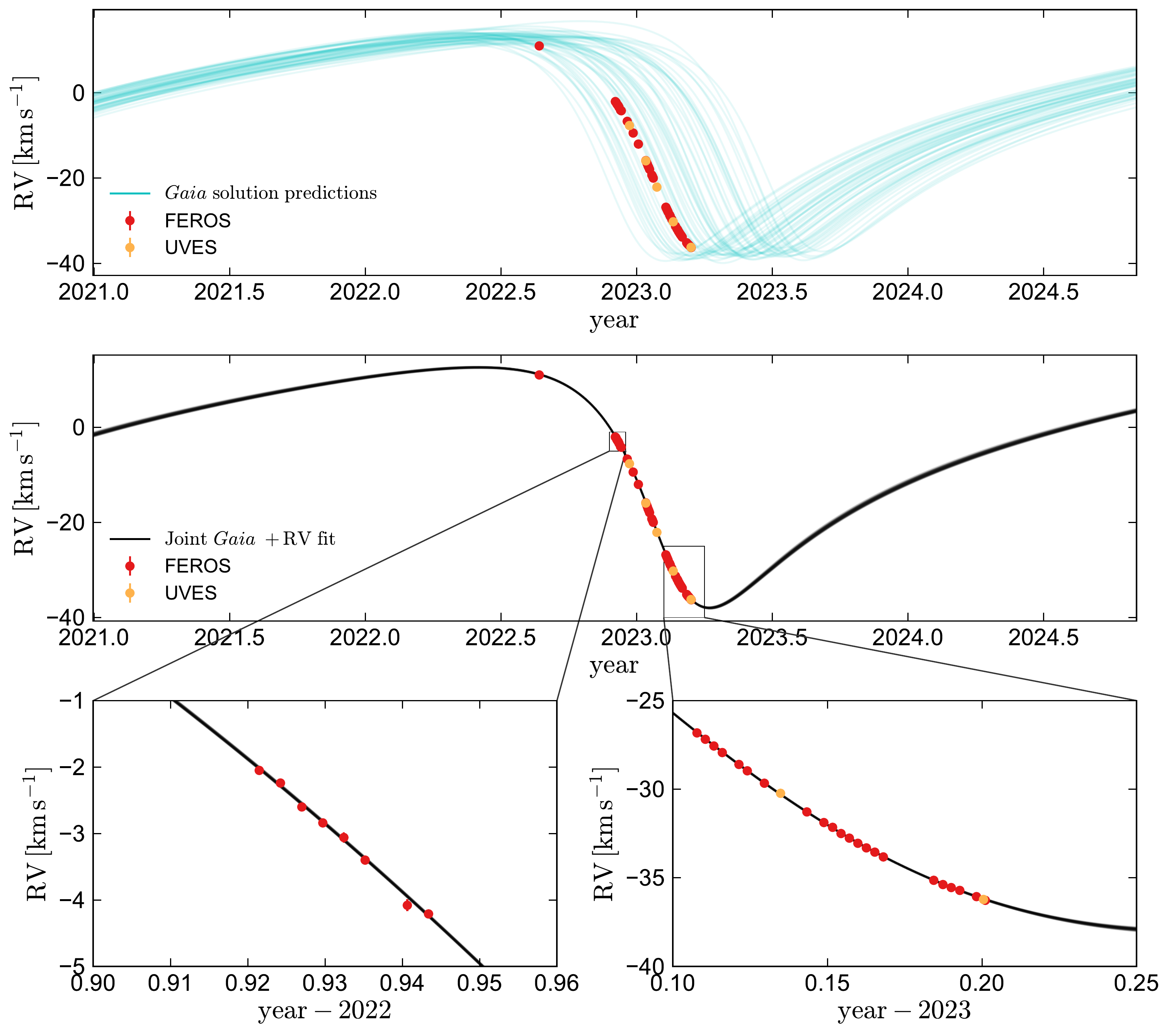}
    \caption{Radial velocities (RVs). In the top panel, we compare predictions of the {\it Gaia} orbital solution (cyan lines) to our measured RVs. There is uncertainty in these predictions, which are based on {\it Gaia} observations taken $\sim 6$ years earlier, but the RVs are consistent with them. In the middle and bottom panels, we show predictions from a joint fit of the RVs and {\it Gaia} constraints. Insets highlight periods with high-cadence data. The RVs have a typical uncertainty of 30\,$\rm m\,s^{-1}$ and span most of the orbit's predicted RV range, so they tightly constrain the orbit when combined with the longer-baseline {\it Gaia} constraints.  }
    \label{fig:rvfig}
\end{figure*}

Figure~\ref{fig:rvfig} (top panel) compares the measured RVs to the predictions of the {\it Gaia}-only solution. There is significant ($\pm 3$ months) uncertainty in the predicted RV turnover time, presumably because the {\it Gaia} data did not cover a full period and were obtained two orbital cycles before our follow-up (see Appendix~\ref{sec:appendix_gost}). The follow-up RVs are fully consistent with the {\it Gaia}-only solution. 

Perhaps counterintuitively, the best-fit period obtained from the {\it Gaia}-only fit is $P_{\rm orb} = 1300\pm 26$ days, which is different at the $\sim 1\sigma$ level from the period constraint in the {\it Gaia} archive, $P_{\rm orb} = 1352\pm 45$ days. Several other parameters are also slightly discrepant. The reason for this difference is that our modeling explicitly links the spectroscopic and astrometric Thiele-Innes parameters via the requirement that $a_0=a_1$, and by the requirement that a single pair of ($\omega, i$) can explain all six Thiele-Innes parameters.  
In contrast, the {\it Gaia} data processing only requires that the spectroscopic and astrometric orbits have the same period, eccentricity, and eccentric anomaly at fixed time; $A,B,F,G,C$, and $H$ are fit to the astrometric and RV time series without enforcing constraints between them \citep[see][]{Pourbaix2022}. Inspection of Equations~\ref{eq:Apred}-\ref{eq:Hpred} reveals that the six Thiele-Innes parameters are functions of four Campbell elements $(a_0, \omega, \Omega, i$), or of five if $a_1$ is fit as a separate parameter. This makes the problem over-constrained.  In the presence of noise, there is thus no guarantee that a single set of orbital parameters can actually match all the Thiele-Innes parameters, even for a perfectly well-behaved orbit. Our approach minimizes the difference between the predicted and measured parameters. As we show in Table~\ref{tab:innes_elememts}, there does indeed exist a single set of orbital parameters that can reproduce all the {\it Gaia} constraints and our follow-up RVs.

\subsubsection{Joint constraints}
The middle and bottom panels of Figure~\ref{fig:rvfig} compare the measured RVs to predictions of the joint {\it Gaia}+RV fit. With the RVs included, the fit includes data taken over the course of $\sim 3$ orbital periods, resulting in a tightly constrained orbit. The follow-up RVs cover most of the best-fit orbit's dynamic range in RV and thus constrain the RV mass function more tightly than the pure-{\it Gaia} constraints (Section~\ref{sec:jointconstraints}). We show the residuals of the measured RVs with respect to the best-fit orbit in the top panel of Figure~\ref{fig:mass_fn}. The fit is good, with $\chi^2/N_{\rm RVs} \approx 0.7$, suggesting that the typical RV uncertainty is slightly overestimated.

\begin{figure*}
    \centering
    \includegraphics[width=0.9\textwidth]{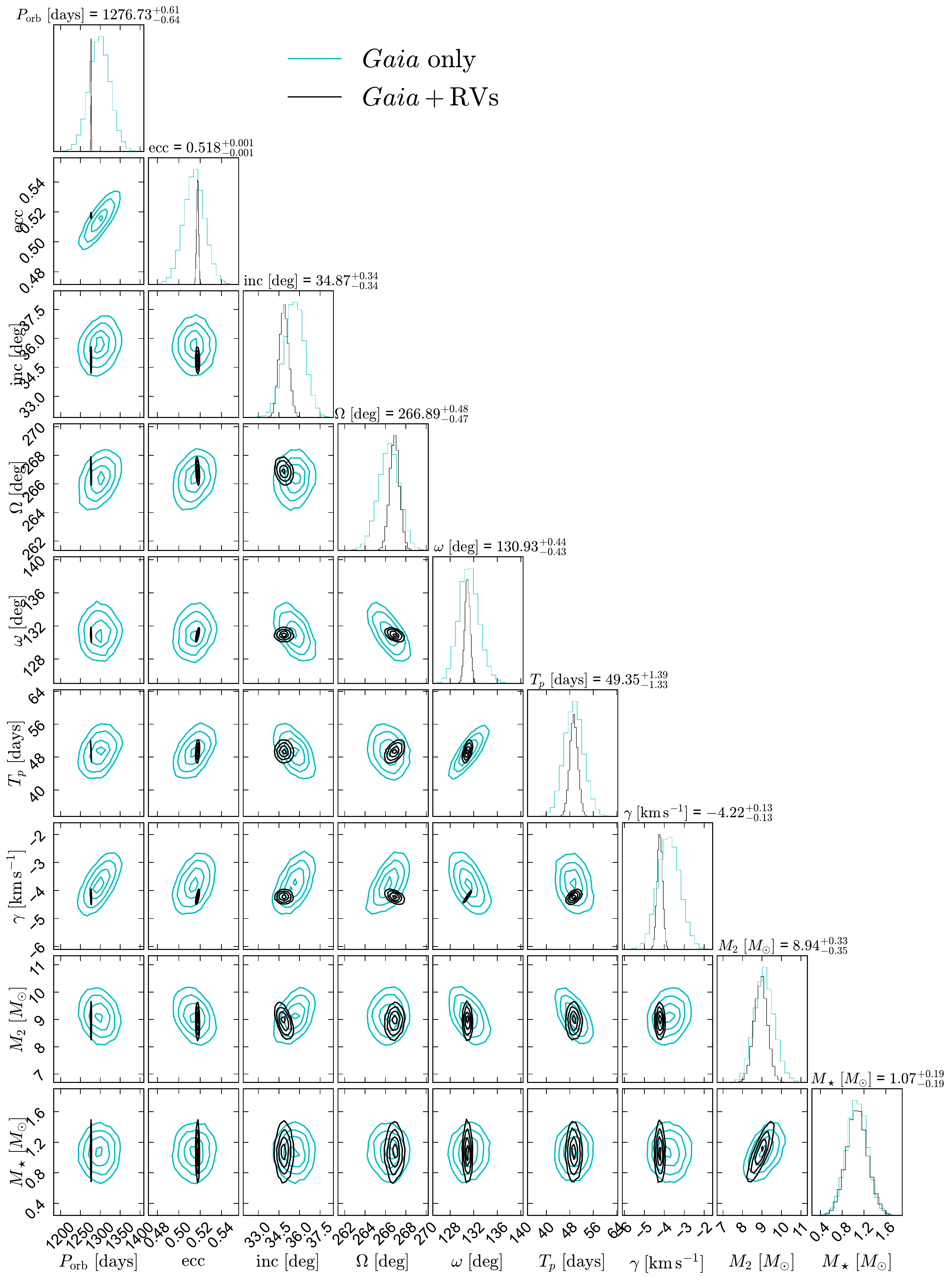}
    \caption{Comparison between constraints obtained from joint fitting of our RVs and the {\it Gaia} data (black) and constraints from {\it Gaia} alone (cyan). The two sets of constraints are generally consistent, but those conditioned on the RVs are tighter. Numbers on the diagonals reflect the joint constraints. Both sets of constraints are listed in Table~\ref{tab:system}. }
    \label{fig:corner_plot_comparison}
\end{figure*}

 In Figure~\ref{fig:corner_plot_comparison}, we compare the constraints from joint modeling to those based only on the {\it Gaia} constraints. Reassuringly, the two sets of constraints are consistent, but the constraints that include our RVs are tighter. 
 
 \subsubsection{Companion mass and inclination}
 \label{sec:jointconstraints}
 Our joint fit constrains the mass of the unseen companion to $M_2 = 8.9\pm 0.3\,M_{\odot}$. The dominant source of uncertainty in this estimate is the uncertainty in the red giant mass.

\begin{figure}
     \centering
     \includegraphics[width=\columnwidth]{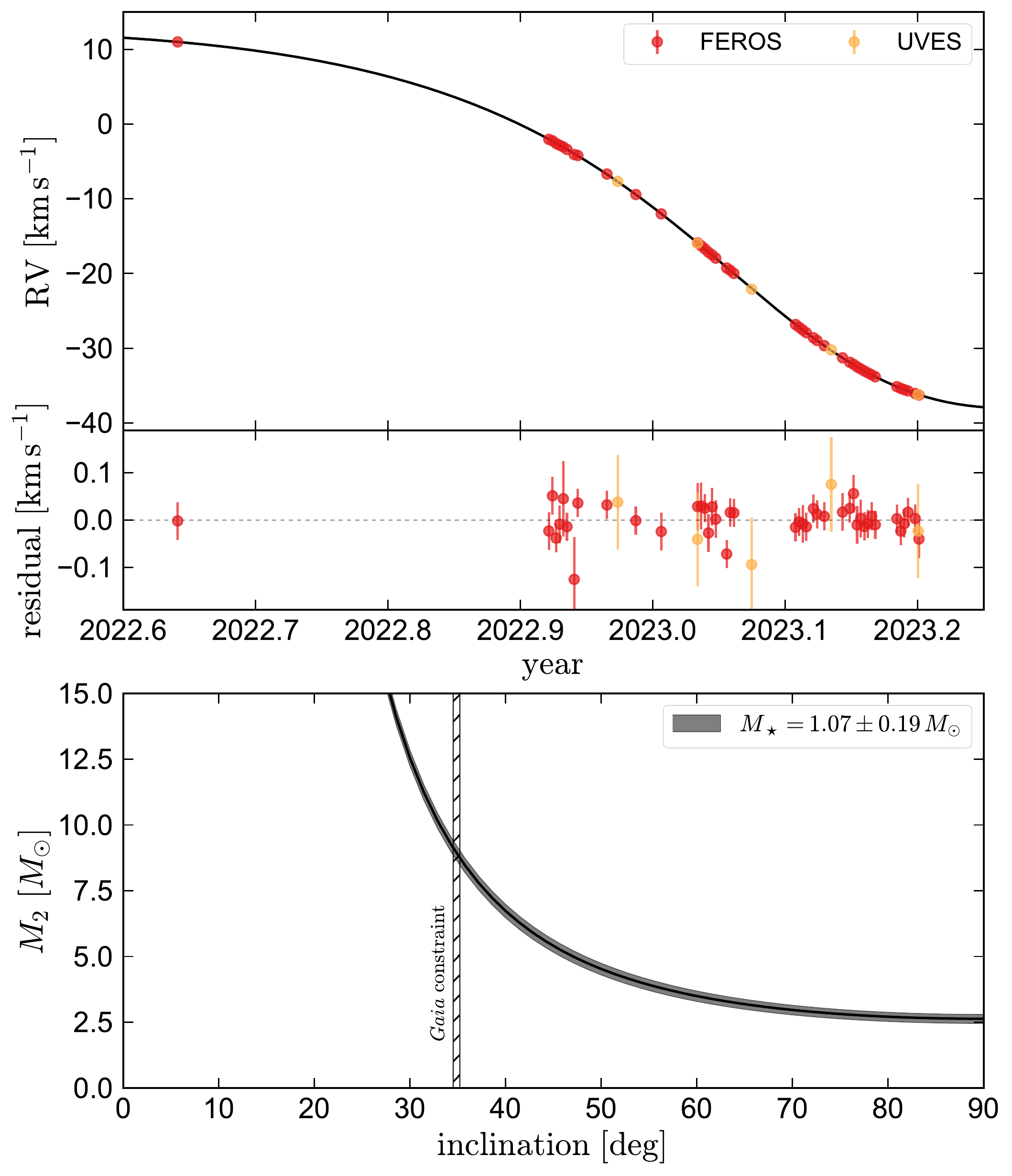}
     \caption{Top: comparison between the best-fit {\it Gaia} + RV solution and the measured RVs. The median absolute residual is 0.02 $\rm km\,s^{-1}$, consistent with the RV uncertainties. Bottom:
     mass of the unseen companion implied by the RV orbital solution as a function of the assumed inclination. Shaded region shows the $\pm 1\,\sigma$ mass range for the luminous star. Hatched vertical band shows the {\it Gaia} inclination constraint, which implies a companion mass of $M_2 \approx 8.9\pm 0.3\,M_{\odot}$. If the orbit were edge-on (in conflict with the astrometric solution), the minimum companion mass would be $\approx 2.6\,M_{\odot}$. }
     \label{fig:mass_fn}
 \end{figure}

 The astrometric constraint on the eccentricity plays a critical role in constraining the mass, because the inclination, $i\approx 35$ deg, is relatively face-on. This means that the component of the star's orbital velocity projected onto our line-of-sight is only $\sin(35\,\rm deg)\approx 57\%$ of the total orbital velocity. We show how the implied companion mass varies with inclination in Figure~\ref{fig:mass_fn}, where we plot the implied companion mass as a function of assumed inclination, giving the RV mass function, $f\left(M_{2}\right)=P_{{\rm orb}}K_{\star}^{3}\left(1-e^{2}\right)^{3/2}/\left(2\pi G\right)\approx1.33\,M_{\odot}$. If the orbit were edge-on, the companion mass implied by the period, RV semi-amplitude, and eccentricity would be $M_2 \approx 2.6\,M_{\odot}$. While this is much lower than the fiducial mass constraint when the inclination measurement is taken into account, it is still too massive for the unseen companion to be any plausible luminous star, whether a single star or a binary itself (Section~\ref{sec:nature_of_companion}).

Only 18\% of randomly oriented orbits are expected to have an inclination as low as 35 degrees.  We note that in contrast to RV surveys, astrometric binary surveys have a selection bias in favor of low inclinations \citep[e.g.][]{Arenou2022}.
The good agreement between the RV prediction from the {\it Gaia} solution and our measured RVs, as well as the fact that RVs alone rule out a stellar companion even for an edge-on orbit (Figure~\ref{fig:seds}), suggests that the {\it Gaia} inclination constraint is reliable.

\subsubsection{Pure RV constraints}
\label{sec:pureRV}
Recognizing that the {\it Gaia} orbital solution could be wrong, we now explore constraints that can be obtained from our follow-up RVs alone. 

\begin{figure}
     \centering
     \includegraphics[width=\columnwidth]{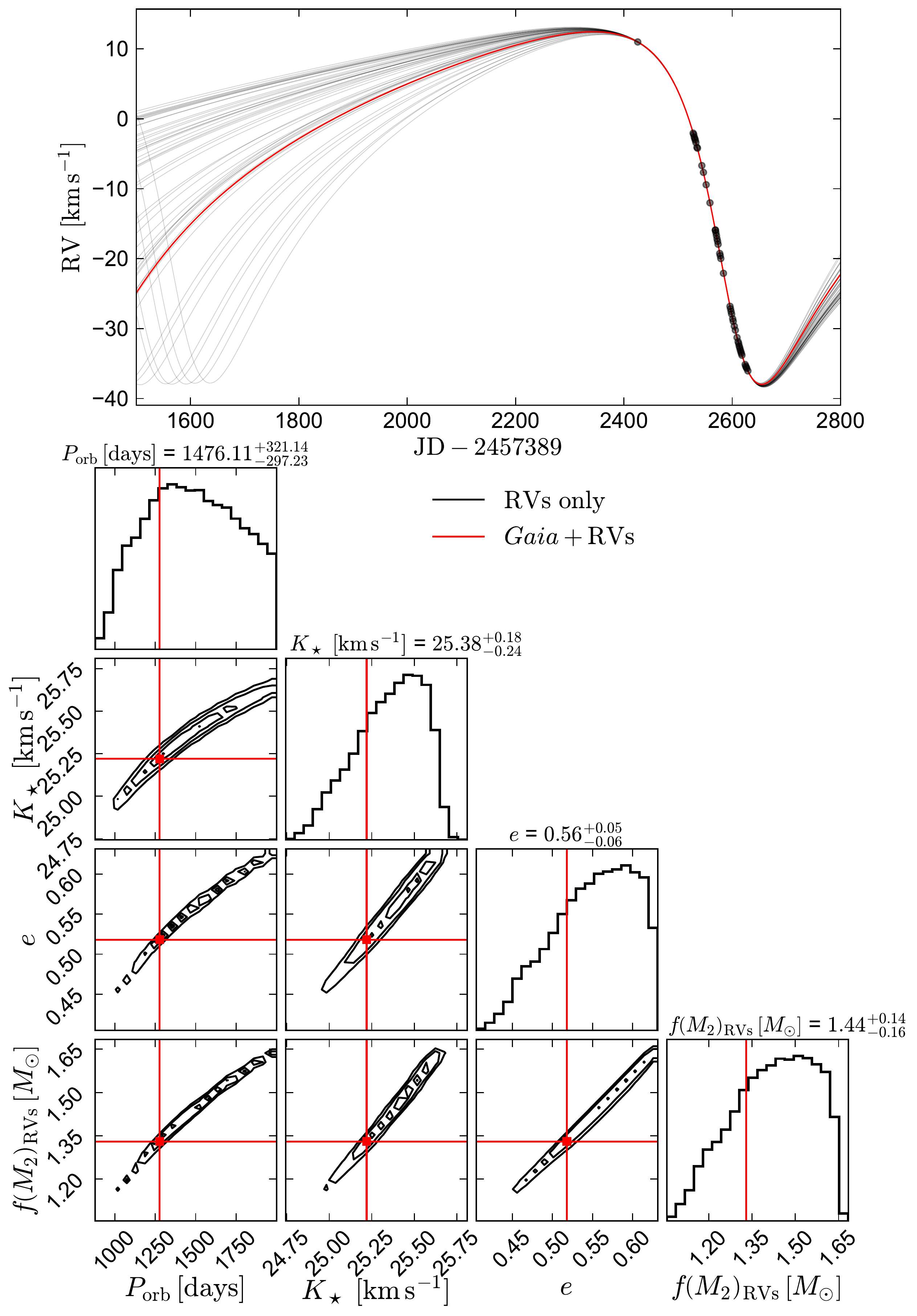}
     \caption{Constraints on the Gaia BH2 orbit from RVs alone, ignoring the {\it Gaia} constraints entirely. Black lines show predicted RV curves (top) and posterior constraints (bottom) from fitting our follow-up RVs. Red lines show the best-fit joint {\it Gaia} + RV solution. The two sets of constraints are fully consistent. Although they cover only a small fraction of the orbit, the RVs alone rule out periods below $\sim 1150$ days and RV mass functions below $\sim 1.25\,M_{\odot}$. The RVs alone do not provide a meaningful {\it upper} limit on the orbital period (here we adopted a prior with an upper limit of $P_{\rm orb} < 2000$ days), but any longer period would imply an even larger companion mass function. }
     \label{fig:rvonly}
 \end{figure}

We fit the FEROS and UVES RVs with \texttt{the Joker} \citep{Price-Whelan2017}, which uses rejection sampling to obtain unbiased posterior samples consistent with a set of RVs. We used a $p\left(P_{{\rm orb}}\right)\propto1/P_{{\rm orb}}$ prior between 100 and 2000 days, and broad uniform priors on other orbital parameters. Even with $10^{10}$ prior samples, \texttt{the Joker} returned only a few samples that were marginally consistent with the RVs, and all of these samples were concentrated in a single posterior mode. This reflects the fact that the RVs strongly constrain the orbit, such that only a tiny fraction of orbital parameter space is consistent with the data. To better sample the posterior, we initialized standard MCMC chains near the maximum-likelihood sample and drew more samples using \texttt{emcee}. The results are reported in the bottom block of Table~\ref{tab:system} and are shown in Figure~\ref{fig:rvonly}. In the top panel, gray lines show RV predictions from random posterior draws; red line shows the prediction of the best-fit joint {\it Gaia} + RV solution. In the bottom panels, contours show constraints from the RVs alone, while the red square marks the best-fit joint solution. 

The RV-only constraints are fully consistent with those from the joint fit. Because our RVs cover only a small fraction of the period, the orbit is not fully constrained by RVs alone. In particular, the RVs allow for orbits with arbitrarily long periods combined with increasingly large eccentricities. For this reason, there are many posterior samples with periods near our adopted upper limit of 2000 days. However, with a prior of $P_{\rm orb} < 2000$ days, the orbit becomes well-constrained.  Most importantly, the RVs alone provide a well-defined {\it minimum} period and mass function, ruling out any $P_{\rm orb} \lesssim 1150$ days and any RV mass function $f(M_2)_{\rm RVs}\lesssim 1.25\,M_{\odot}$. This means that -- independent of the {\it Gaia} data -- the RV orbit and luminous star mass imply a companion mass of at least $2.5\,M_{\odot}$. 

\begin{table*}
\begin{tabular}{lllll}
Parameter & units & {\it Gaia} solution  & Joint RVs+astrometry constraint & How many sigma discrepant? \\
\hline
\texttt{a\_thiele\_innes} & mas & $2.48\pm 0.05$ & $2.43 \pm 0.02$ & 0.7 \\
\texttt{b\_thiele\_innes} & mas & $2.45\pm 0.10$  & $2.31 \pm 0.02$   & 1.2 \\
\texttt{f\_thiele\_innes} & mas & $-1.98\pm 0.10$  & $-1.84 \pm 0.03$   & 1.1 \\
\texttt{g\_thiele\_innes} & mas & $3.01\pm 0.11$  & $2.91 \pm 0.02$  & 0.8 \\
\texttt{c\_thiele\_innes} & mas & $2.06\pm 0.11$  & $1.914 \pm 0.014$  & 1.2 \\ 
\texttt{h\_thiele\_innes} & mas & $-1.95\pm 0.12$  & $-1.66\pm 0.014$  & 2.2 \\ 
\texttt{period}         & days& $1352 \pm 45$ &$1277 \pm 1$ &  1.6 \\ 
\texttt{eccentricity}   &     & $0.530\pm 0.015$  & $0.518\pm 0.001$ & 0.8 \\
\texttt{t\_periastron}   & days& $52.7\pm 3.9$  & $49.3\pm 1.4$  & 0.6 \\ 
\texttt{center\_of\_mass\_velocity}   & $\rm km\,s^{-1}$ & $-3.2\pm 0.6$  & $-4.2\pm 0.1$  & 1.4 \\ 
\end{tabular}
\caption{Comparison of the {\it Gaia} constraints on Gaia BH2's orbit to constraints from the joint RVs+astrometry fit. All parameters are consistent at the $\sim$2$\sigma$ level. The most discrepant parameter is \texttt{h\_thiele\_innes}, for which there is 2.2$\sigma$ tension. Overall, this comparison suggests that the {\it Gaia} solution is reliable, though its uncertainties could be underestimated slightly.}
\label{tab:innes_elememts}
\end{table*}

\subsection{Spectral analysis}
\label{sec:abundances}
We analyzed the UVES spectra to measure atmospheric parameters and abundances of the red giant and place further constraints on the light contributions of the secondary, using a combination of ab-initio and empirical modeling. Most of our analysis focused on the spectrum obtained on JD 2459935, but we verified that there are no significant differences between the rest-frame UVES spectra taken on different nights. 

We first fit the UVES spectra using the Brussels Automatic Code for Characterizing High accUracy Spectra \citep[BACCHUS;][]{Masseron2016} with the same set up as in \citet{El-Badry2023}. BACCHUS enables us to derive the stellar atmospheric parameters, including the effective temperature ($T_{\rm eff}$), surface gravity (log$g$), metallicity ([Fe/H]) and microturblent velocity ($v_{\rm micro}$) by assuming Fe excitation/ionization balance; i.e., the requirement that lines with different excitation potentials all imply the same abundances. We use the fifth version of the Gaia-ESO atomic linelist \citep{Heiter2021}.  Hyperfine structure splitting is included for Sc I, V I Mn I, Co I, Cu I, Ba II, Eu II, La II, Pr II, Nd II, Sm II \citep[see more details in][]{Heiter2021}. We also include molecular line lists for the following species: CH \citep{Masseron2014}, and CN, NH, OH, MgH and  C$_{2}$ (T. Masseron, private communication). Finally, we also include the SiH molecular line list from the Kurucz linelists\footnote{http://kurucz.harvard.edu/linelists/linesmol/}. Spectral synthesis for BACCHUS is done using the TURBOSPECTRUM \citep{Alvarez1998, Plez2012} code along with the line lists listed above and the MARCS model atmosphere grid \citep{Gustafsson2008}. 

Once the stellar parameters were determined, the model atmosphere is fixed and individual abundances were derived using BACCHUS' `abund' module. For each spectral absorption feature, this module creates a set of synthetic spectra, which range between -0.6 $<$ [X/Fe] $<$ +0.6~dex, and performs a $\chi^2$ minimization between the observed and synthetic spectra. The reported atmospheric [X/Fe] abundances are the median of derived [X/Fe] across all lines for a given species. The uncertainty in the atmospheric [X/Fe] is defined as the dispersion of the [X/H] abundance measured across all lines for a species. If only 1 absorption line is used, we conservatively assume a [X/Fe] uncertainty of 0.10~dex. For a more detailed discussion of the BACCHUS code, we refer the reader to Section~3 of both \cite{Hawkins2019, Hawkins2020}. We ran BACCHUS on the UVES spectrum after merging individual de-blazed orders and performing a preliminary continuum normalization using a polynomial spline. Further normalization is performed by BACCHUS. Spectral regions affected by telluric absorption were masked during fitting, and we only fit the portion of the spectrum with $\lambda < 7600\,$\AA, where tellurics are less severe. 

The resulting stellar parameters and abundances are listed in Table~\ref{tab:bacchus}. Figure~\ref{fig:modelspec} compares portions of the observed spectrum to the best-fit BACCHUS model. The fit is overall quite good: while there are some deviations between the model and data, the quality of the fit is typical for ab-initio (as opposed to data-driven) fits to high-resolution spectra. The remaining differences between data and model can be attributed to a combination of imperfect continuum normalization and imperfections in the line list.  

The most important conclusion from this analysis is that the star's iron abundance is subsolar ($[\rm Fe/H] = -0.22$), but the abundances of $\alpha-$elements (i.e.; O, Mg, Si, Ca, and Ti) are enhanced relative to solar, with $[\rm \alpha/Fe]=0.26\pm 0.05$. This abundance pattern is characteristic of old stars in the Galactic thick disk or metal-rich halo, which formed from material that was enriched by type II SNe, with less enhancement of iron-peak elements by SNe Ia than typical stars in the thin disk. If Gaia BH2 formed in the thick disk, this would imply an age of 8-12 Gyr \citep[e.g][]{Nissen2020, Xiang2022} and (assuming mass transfer was negligible) a stellar mass of $M_\star \approx 1.0 \pm 0.07\,M_{\odot}$ for the giant.

Beyond the strong $\alpha-$ enhancement, the abundances of the giant appear normal. We compared the star's position in the [X/Fe] vs. [Fe/H] plane to the population of stars in the solar neighborhood with abundances measured by \citet{Adibekyan2012}, \citet{Bensby2014}, and \citet{Battistini2016}, finding that all elements fall within the $\sim 2\,\sigma$ scatter of the observed local population. Lithium is not detected, as expected for a star on the giant branch that has undergone first dredge-up. There is no evidence of enhancement in neutron-capture elements relative to typical thick-disk stars.

\subsubsection{Empirical comparison to other red giant spectra and constraints on luminous companions}
\label{sec:uves_luminous}
\begin{figure*}
    \centering
    \includegraphics[width=\textwidth]{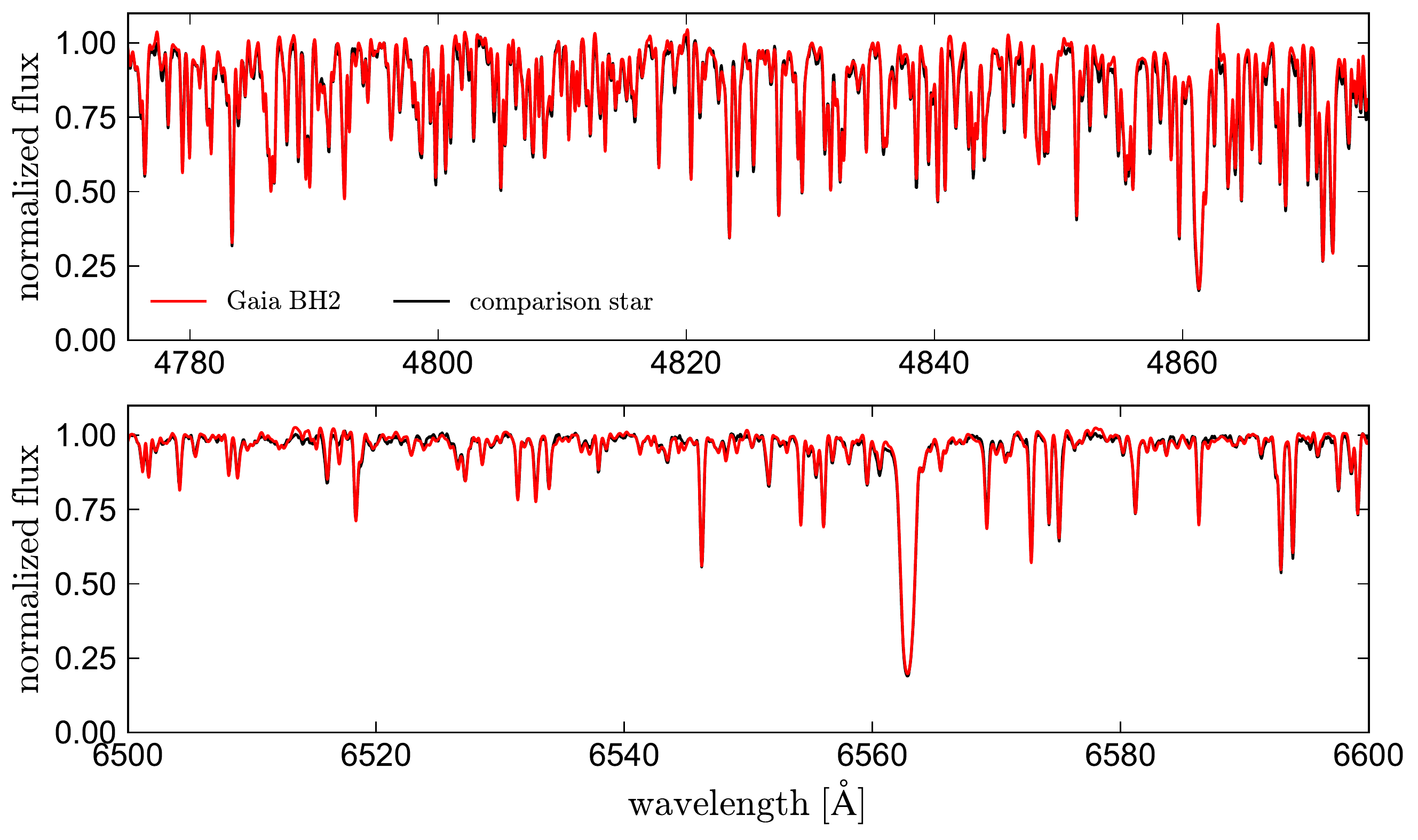}
    \caption{Spectral cutouts of Gaia BH2 (red) compared to a reference star (black), a thick-disk star with similar stellar parameters and abundances observed by the GALAH survey. The two spectra are very similar. This rules out significant light contributions from a companion and speaks against ongoing mass transfer. It also indicates that the abundance pattern of the giant is not grossly unusual.  }
    \label{fig:galah}
\end{figure*}

To search for possible anomalous spectral features -- which could arise, for example, due to a luminous companion, emission from an accretion disk, or absorption by an accretion stream -- we compared the spectrum of Gaia BH2 to spectra of stars with similar stellar parameters and abundances observed by the GALAH survey \citep{DeSilva2015, Buder2021}. To this end, we degraded the observed UVES spectrum to $R=28,000$, shifted it to rest frame, and identified its nearest neighbor in pixel space among stars observed by GALAH with SNR > 100. The closest match was {\it Gaia} DR3 source 5909513401713377280, which in GALAH DR3 has $T_{\rm eff} = 4591\pm 74\,\rm K$, $\log g = 2.45\pm 0.19$, $\rm [Fe/H]=-0.36\pm 0.05$, and $[\rm \alpha/Fe]=0.28\pm 0.01$. These parameters are all similar to those we find for Gaia  BH2, providing independent validation of our BACCHUS-inferred parameters. In Figure~\ref{fig:galah}, we compare the normalized and resolution-matched spectra of the two sources in the 1st and 3rd GALAH wavelength windows, which respectively contain H$\beta$ and H$\alpha$. The spectra are nearly indistinguishable. 

\begin{figure*}
    \centering
    \includegraphics[width=\textwidth]{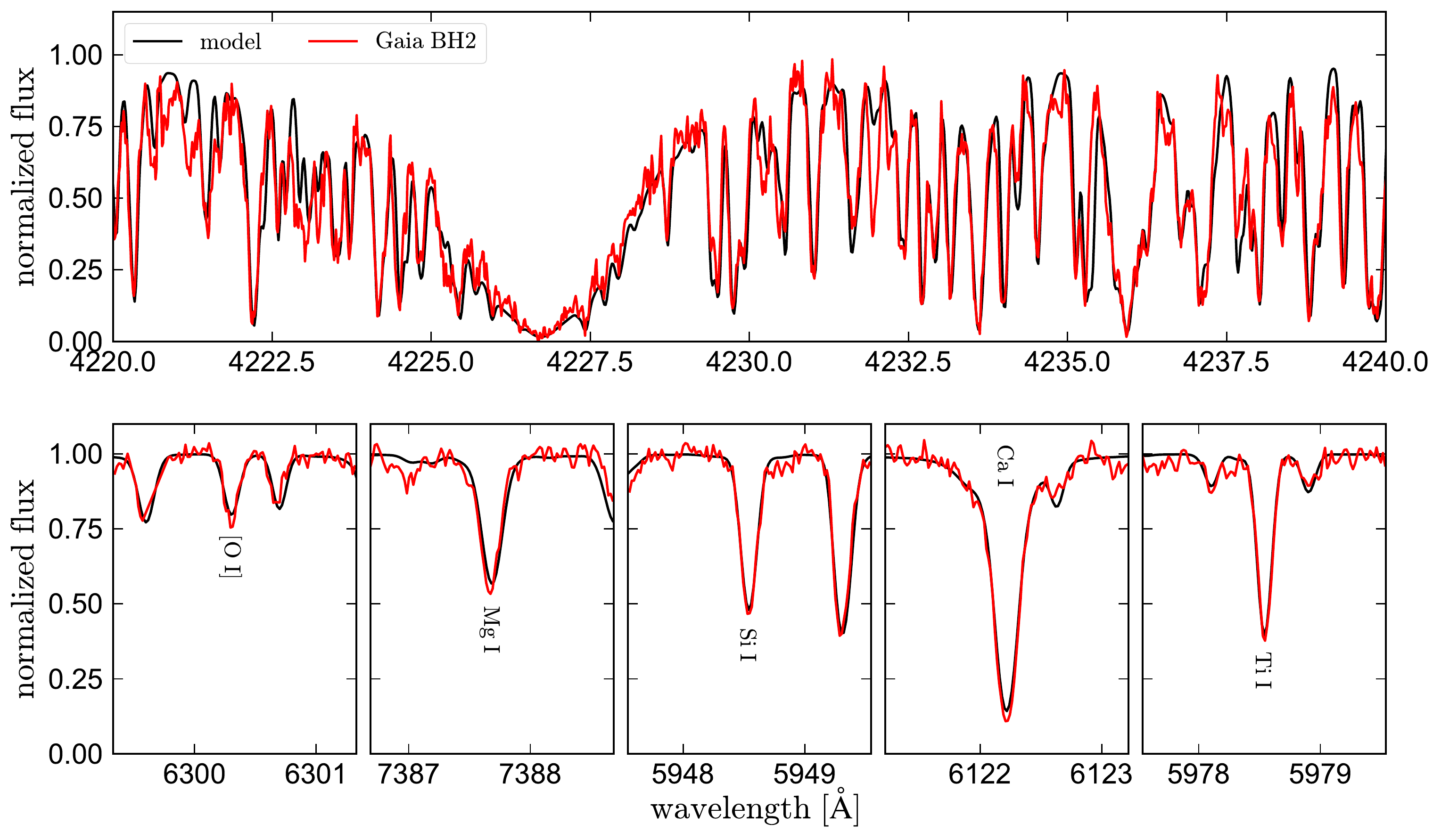}
    \caption{UVES spectral cutouts of Gaia BH2 (red) compared to BACCHUS spectral model (black; parameters are in Table~\ref{tab:bacchus}). Top panel shows a 20\,\AA-wide region containing the Ca I resonance line at 4227\,\AA, as well as many other metal lines. Bottom panels highlight lines from 5 different $\alpha$-elements, whose abundances are all enhanced relative to the solar abundance pattern. The deepest metal lines reach a depth of $\approx 2\%$ of the continuum; this rules out luminous companions that contribute more than $\approx 2\%$ of the light at these wavelengths.  }
    \label{fig:modelspec}
\end{figure*}

The similarity of the two observed spectra, and the good agreement between the data and the BACCHUS model spectrum in Figure~\ref{fig:modelspec}, speaks against the presence of a luminous secondary. The quantitative upper limit depends on the spectral type of the putative companion: a rapidly rotating secondary would be harder to detect. However, the depth of the observed absorption lines rules out even a companion that contributes only continuum. The total flux in the cores of the deepest lines is only $\approx 2\%$ of the continuum value (Figure~\ref{fig:modelspec}). A companion contributing more than $\approx 2\%$ of the total light at 4000-4300\,\AA\, would dilute the absorption lines beyond this value and is thus ruled out. We conservatively adopt a $< 3\%$ flux limit on the total flux; this is the origin of the blue upper limit in Figure~\ref{fig:seds}.

\subsection{Galactic orbit}
\label{sec:galpy} 

\begin{table}
\begin{tabular}{lllll}
Parameter & BACCHUS Constraint & $N_{\rm lines}$ \\
\hline
$T_{\rm eff}\,[\rm K]$ & $4604 \pm 87$  &   \\
$\log\left(g/{\rm cm\,s^{-2}}\right)$ & $ 2.71 \pm 0.24$  &   \\
$v_{\rm micro}\,\left[{\rm km\,s^{-1}}\right]$ & $1.16\pm 0.04$ &  \\
$\rm [Fe/H]$& $-0.22 \pm 0.02$ &    \\
$\rm [\alpha/Fe]$& $0.26 \pm  0.05$ &    \\
$\rm [Na/Fe]$& $0.19 \pm 0.03$ & 7  \\
$\rm [Mg/Fe]$& $0.32 \pm 0.15$ & 4  \\
$\rm [Al/Fe]$& $0.51 \pm  0.1$ & 1  \\
$\rm [Si/Fe]$& $0.21 \pm 0.05$& 5  \\
$\rm [Ca/Fe]$& $0.2 \pm 0.02$ & 14  \\
$\rm [O/Fe]$& $0.32 \pm 0.1$ & 1  \\
$\rm [Ti/Fe]$& $0.28 \pm 0.02$ & 31  \\
$\rm [V/Fe]$& $0.05 \pm 0.03$ & 24  \\
$\rm [Sc/Fe]$& $0.20 \pm 0.05$ & 4  \\
$\rm [Cr/Fe]$& $-0.14 \pm 0.05$ & 11  \\
$\rm [Mn/Fe]$& $-0.33 \pm 0.05$ & 3  \\
$\rm [Co/Fe]$& $0.12 \pm 0.04$ & 10  \\
$\rm [Ni/Fe]$& $0.06 \pm  0.03$ & 16  \\
$\rm [Cu/Fe]$& $-0.21 \pm  0.01$ & 2  \\
$\rm [Zn/Fe]$& $-0.02 \pm 0.08$ & 4  \\
$\rm [Sr/Fe]$& $0.11 \pm  0.04$ & 4  \\
$\rm [Y/Fe]$& $-0.30 \pm 0.10$ & 8  \\
$\rm [Zr/Fe]$& $0.01 \pm 0.03$ & 9  \\
$\rm [Ba/Fe]$& $0.00 \pm 0.06$ & 5  \\
$\rm [La/Fe]$& $0.14 \pm 0.05$ & 8  \\
$\rm [Nd/Fe]$& $0.05 \pm  0.07$ & 7   \\

\end{tabular}
\caption{Parameters of the red giant inferred by BACCHUS. $N_{\rm lines}$ is the number of independent absorption lines used to infer each element. $\alpha$ is an average of O, Mg, Si, Ca, and Ti.}
\label{tab:bacchus}
\end{table}

\begin{figure*}
    \centering
    \includegraphics[width=\textwidth]{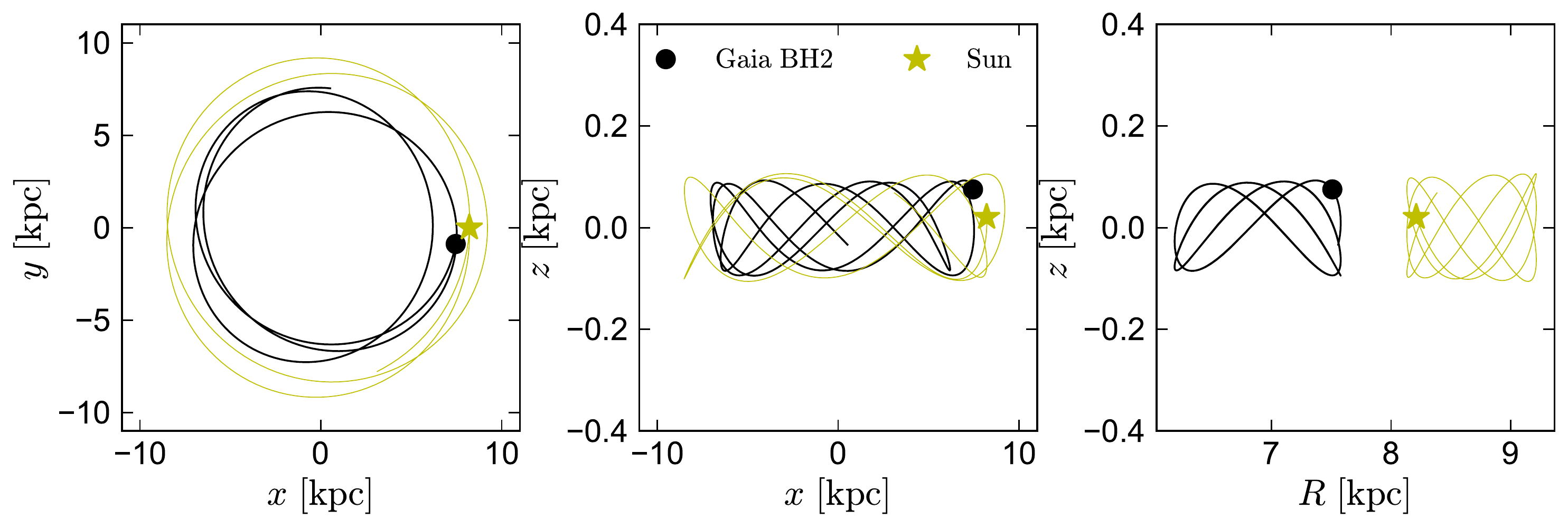}
    \caption{Galactic orbit of Gaia BH2, calculated backward for 500 Myr from the measured proper motion and center-of-mass RV. For comparison, we show the Sun's orbit calculated over the same period. The orbit is typical of a thin-disk star. This suggests that the BH was not born with a large natal kick and also likely did not form in a classical globular cluster.}
    \label{fig:galpy}
\end{figure*}

To investigate the Galactic orbit of Gaia BH2, we used its parallax and proper motion from the {\it Gaia} astrometric binary solution, as well as the center-of-mass RV inferred from the joint fit, as starting points to compute its orbit backward in time for 500 Myr using \texttt{galpy} \citep[][]{Bovy2015}. We used the Milky Way potential from \citet{McMillan2017}. The result is shown in Figure~\ref{fig:galpy}; for comparison, we also show the orbit of the Sun. The orbit is typical of a {\it thin}-disk star, with modest eccentricity and excursions above the disk midplane limited to $\pm 80$ pc. 

\subsubsection{Interpretation of the abundances and kinematics}
\label{sec:abund_interp}
\begin{figure*}
    \centering
    \includegraphics[width=\textwidth]{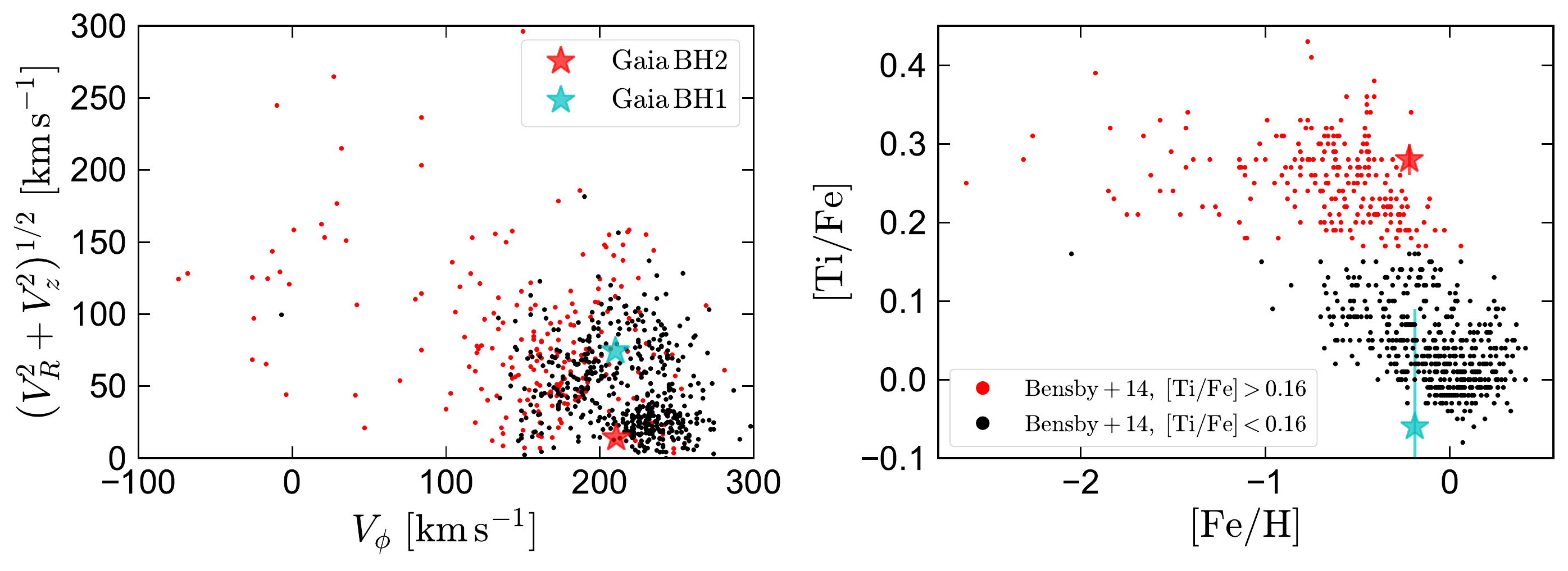}
    \caption{Toomre diagram (left) and [Ti/Fe] vs. [Fe/H] diagram (right; Ti is a represenative $\alpha$-element). Small points show stars in the solar neighborhood from \citet{Bensby2014} that are $\alpha-$rich (red) and $\alpha-$poor (black).  Large red and cyan stars show Gaia BH2 and BH1, respectively. Gaia BH2 is kinematically part of the thin disk. Its $\alpha-$enhanced abundances are, however, more characteristic of the thick disk. The opposite is true for Gaia BH1. }
    \label{fig:alphafe}
\end{figure*}

The combination of thin-disk kinematics with $\alpha-$rich chemistry is unusual. This is illustrated in Figure~\ref{fig:alphafe}, where we compare Gaia BH2's chemistry and kinematics to the solar neighborhood sample from \citet{Bensby2014}. The left panel shows the Toomre diagram, while the right panel shows the abundance of titanium (a representative $\alpha-$element) relative to iron. 
Most of the $\alpha-$enhanced stars in the \citet{Bensby2014} sample have thick-disk kinematics. In contrast, Gaia BH2 is strongly enhanced in all $\alpha$ elements but has a Galactic orbit typical of the thin disk. 

Given its thin-disk-like orbit, it is tempting to attribute the $\alpha-$enhancement of Gaia BH2 to pollution from the BH progenitor. Indeed, similar $\alpha-$enhancement in the donor stars of BH X-ray binaries has been interpreted as evidence for pollution during the supernova event. For example, 0.2-0.5 dex enhancement of some $\alpha-$elements in the donor stars in GRO J1655-40 and V404 Cyg -- the BH X-ray binaries whose detailed abundance patterns can be measured most readily -- has been interpreted by \citet{Israelian1999} and \citet{GonzalezHernandez2011} as evidence that the BHs formed in a supernova. In those systems, the observed abundance patterns can plausibly be explained by pollution from a supernova with yields similar to those predicted for stars with initial masses of $25-40\,M_{\odot}$.

It is more challenging to apply a similar a similar explanation to Gaia BH2. 
At the periastron distance of $\approx 2.4$\,au, a star with $R_\star \approx 1\,R_{\odot}$ subtends only $\approx 10^{-6}$ of the sky as seen from the BH progenitor. The supernova ejecta are expected to consist mostly ($\approx 70\%$) of oxygen \citep{Thielemann1996} and must have had a total mass of $\lesssim 10\,M_{\odot}$ in order for the binary to remain bound. For a spherically symmetric explosion, we thus expect at most $7\times 10^{-6}\,M_{\odot}$ of oxygen to have been deposited on the companion during a supernova. During first dredge-up, this would have been mixed with the $\approx 0.005\,M_{\odot}$ of oxygen already present in the star's convective envelope, and would not significantly increase the surface oxygen abundance observable today. Similar calculations apply to the predicted yields of other $\alpha-$elements. It thus seems unlikely that the observed $\alpha-$enhancement is a consequence of pollution by supernova ejecta, if these ejecta escaped the BH progenitor at high velocity and were deposited on the secondary directly.

It is possible, however, that some of the ejecta escaped  the BH progenitor at low velocity, reached a separation of several au, and remained bound to the binary. In this case, a significant amount of the material may have eventually been accreted by the companion star, and it is possible that this supplied the $\approx 0.005\,M_{\odot}$ of $\alpha-$elements required to explain the observed abundances. The expected total accretion onto the secondary in this scenario will depend on the energetics of the explosion and on the structure of the BH progenitor before its death; detailed calculations are required to explore this further.

\subsection{X-ray observations}
\label{sec:chandra}
We observed Gaia BH2 for 20ks using the Advanced CCD Imaging Spectrometer \citep[ACIS;][]{Garmire2003} on board the {\it Chandra} X-ray telescope on 2023 January 25 (proposal ID 23208881; PI: El-Badry). We used the ACIS-S configuration, with a spatial resolution of about 1 arcsec. The observations were timed to occur near the periastron passage, when the separation between the BH and the star was $\approx 2.47\,\rm au$. 

The {\it Chandra} data are shown in the upper left panel of Figure~\ref{fig:xray_radio}. There is no obvious X-ray source coincident with the optical source, whose position is marked with a red circle. We used the Chandra Interactive Analysis of Observations \citep{Fruscione2006} software to place upper limits on the X-ray flux. We first ran the \texttt{chandra\_repro} tool to reprocess the observation; this creates a new bad pixel file and de-streaks the event file. We performed aperture photometry at the optical position of Gaia BH2 in the reprocessed event file using the \texttt{srcflux} tool. We detect no significant flux at the location of Gaia BH2 and obtain a background count rate of $1.45 \times 10^{-4}\,\rm counts\,s^{-1}$. We assume a foreground hydrogen column density of $N_H \approx 1.36\times 10^{21} \textrm{ cm}^{-2}$ based on the optical extinction \citep[e.g.][]{Guver2009} and use the Portable, Interactive Multi-Mission Simulator (PIMMS) tool to calculate the corresponding upper limit on the unabsorbed source flux. Based on the background count rate and $N_H$, and assuming a power law spectrum with photon index 2, we obtain a $2\sigma$ flux limit in the 0.5-7 keV energy range of $F_X < 5.1 \times 10^{-15} \textrm{erg s}^{-1}\textrm{ cm}^{-2}$. This limit is shown as a horizontal dashed line in Figure~\ref{fig:xray_radio}.

\subsection{Radio observations}
\label{sec:meerkat}

We observed Gaia BH2 for 4 hours with the MeerKAT radio telescope in L-band (1.28 GHz) on  2023 January 13 (DDT-20230103-YC-01, PI: Cendes), when the separation between the BH and the star was $\approx 2.54\,\rm au$.  We used the flux calibrator J1939-6342 and the gain calibrator J1424-4913, and used the calibrated images obtained via the SARAO Science Data Processor (SDP)\footnote{https://skaafrica.atlassian.net/wiki/spaces/ESDKB/pages/338723406/}  for our analysis.

The MeerKAT data are shown in the bottom left panel of Figure~\ref{fig:xray_radio}. There is no detectable radio source coincident with Gaia BH2. We measured the flux density using the {\tt imtool} package within {\tt pwkit} \citep{Williams17} at the location of Gaia BH2. The RMS at the source's position is 17 $\mu$Jy, so we report a 2$\sigma$ (3$\sigma$) nondetection of $<34 \mu$Jy ($<51 \mu$Jy). The $2\sigma$ limit is shown with a horizontal dashed line in Figure~\ref{fig:xray_radio}.

\subsection{Expected X-ray and radio flux}
\label{sec:x-ray_expect}
\subsubsection{Bondi-Hoyle-Lyttleton accretion}
 A rough estimate of the expected accretion rate onto the BH can be obtained under the assumption that a spherically-symmetric wind from the giant is accreted at the Bondi-Hoyle-Lyttleton (BHL) rate:
\begin{equation}
\begin{split}
\label{eq:mdot_bhl}
\dot{M}_{{\rm BHL}}	&=\frac{G^{2}M_{{\rm BH}}^{2}\dot{M}_{{\rm wind}}}{v_{{\rm wind}}^{4}d_{\rm p}^{2}} \\
	&=2\times10^{-13}\,M_{\odot}\,{\rm yr}^{-1}\left(\frac{M_{{\rm BH}}}{10\,M_{\odot}}\right)^{2}\left(\frac{\dot{M}_{{\rm wind}}}{10^{-11}M_{\odot}\,{\rm yr}^{-1}}\right)\times\\ &\left(\frac{v_{{\rm wind}}}{150\,{\rm km\,s^{-1}}}\right)^{-4}\left(\frac{d_{\rm p}}{2.5\,{\rm au}}\right)^{-2}.
\end{split}
\end{equation}
The giant's mass-loss rate, $\dot{M}_{\rm wind}$, is uncertain. We consider the prescription from \citet{Reimers1975}:

\begin{equation}
    \label{eq:mdot_reimers}
    \dot{M}_{{\rm wind}}=4\times 10^{-13}\beta_{R}M_{\odot}\,{\rm yr}^{-1}\left(\frac{L_\star}{L_{\odot}}\right)\left(\frac{R_\star}{R_{\odot}}\right)\left(\frac{M_\star}{M_{\odot}}\right)^{-1},
\end{equation}
where $\beta_R$ is a dimensionless constant; we adopt $\beta_R=0.1$ following \citet{Choi2016}. For the wind velocity, a reasonable approximation is that it scales with the escape velocity at the stellar surface, 
\begin{equation}
    \label{eq:vwind}
    v_{{\rm wind}}=600\,{\rm km\,s^{-1}}\beta_{\rm wind}\left(\frac{M_\star}{M_{\odot}}\right)^{1/2}\left(\frac{R_\star}{R_{\odot}}\right)^{-1/2},
\end{equation}
where $\beta_{\rm wind}$ is another dimensionless constant. 

We then assume that the X-ray flux is emitted isotropically with a radiative efficiency $\eta_X$, such that $F_{\rm X,BHL}=\frac{\eta_{X}\dot{M}_{{\rm BHL}}c^{2}}{4\pi d^{2}}$. This leads to a predicted X-ray flux for accretion at the BHL rate:

\begin{equation}
\label{eq:Fx}
\begin{split}
F_{\rm X, BHL}	&=10^{-14}\,{\rm erg\,s^{-1}\,cm^{-2}} \left(\frac{L_\star}{25L_{\odot}}\right)\left(\frac{R_\star}{8\,R_{\odot}}\right)^{3}\left(\frac{M_\star}{M_{\odot}}\right)^{-3}\left(\frac{M_{{\rm BH}}}{9\,M_{\odot}}\right)^{2}\times \\ &\left(\frac{d_{{\rm p}}}{2.5\,{\rm au}}\right)^{-2}\left(\frac{d}{1\,{\rm kpc}}\right)^{-2}\left(\frac{\eta_{X}}{10^{-4}}\right)\left(\frac{\beta_{R}}{0.1}\right)\left(\frac{\beta_{{\rm w}}}{1}\right)^{4},
\end{split}
\end{equation}
where we have scaled to parameters similar to Gaia BH2.
Given the many approximations that enter this scaling, it is only expected to hold to order of magnitude. The most important prediction of Equation~\ref{eq:Fx} is the strong scaling with the radius and luminosity of the luminous star, $F_{\rm X,BHL} \propto R_\star^3 L_\star$. This means that at fixed mass, a giant leads to a much larger predicted X-ray flux than a dwarf. Physically, this occurs because (a) the giant has a significantly higher mass loss rate, $\dot{M}_{\rm wind}$, and (b), the wind from a giant is slower, allowing the BH to capture a larger fraction of it. As a result, the predicted X-ray flux from Gaia BH2 is $\approx 1,000$ times larger than that from Gaia BH1, even though Gaia BH2 is farther away and has a wider orbit. 

We can also predict the radio flux assuming the source falls on the canonical X-ray -- radio flux correlation. We take the version of this relation from \citet{Gallo2006}: 

\begin{equation}
    \label{eq:LR}
    \log L_{R}=0.6\log L_{X}+0.8\log\left(M_{{\rm BH}}/M_{\odot}\right)+7.3,
\end{equation}
where $L_R$ and $L_X$ represent the radio and X-ray luminosity. Following \citet{Gallo2006}, we translate this into a radio flux density under the assumption of a flat radio spectrum ($S_\nu \propto \nu^0$) up to a maximum frequency $\nu_{\rm max} = 8.4\,\rm GHz$. 

We show the expected X-ray and radio fluxes from Gaia BH2 in Figure~\ref{fig:xray_radio} with red-dashed lines. For each value of $\dot{M}_{\rm BHL}$, we predict the radiative efficiency using the fitting function from \citet{Xie2012}, who used simulations of hot accretion flows to calculate these efficiencies as a function of the accretion rate at the event horizon in Eddington units, for several values of the electron heating parameter, $\delta$. We consider their predictions for $\delta = 0.5$.
For a plausible wind speed of $v_{\rm wind}\approx 150\,\rm km\,s^{-1}$ (somewhat lower than the giant's escape velocity, since the wind must escape the giant's gravitational potential before reaching the BH), the predicted X-ray and radio fluxes are in tension with the observed non-detections. 

\subsubsection{Reduction in $\dot{M}$ near the BH}
\label{sec:wind_losses}
The calculations above implicitly assume that the accretion rate near the event horizon is equal to the BHL rate. This may not be realized in practice if, for example,  a significant fraction of the inflowing material is lost to winds \citep[e.g.][]{Blandford1999}. 
Numerous simulations of radiatively inefficient accretion flows predict that $\dot{M}$ decreases toward the event horizon roughly as $\dot{M} \propto r^{0.5}$, where $r$ is the distance from the BH \citep{Pen2003, Pang2011, Yuan2012, Ressler2021}. There are several proposed physical explanations for this scaling. For concreteness, we consider the model from \citet{Yuan2012}, in which $\dot{M} \propto r^s$, where $s\approx 0.5$ down to $r \sim 10 R_s$, and $s\approx 0$ at smaller radii, where $R_s$ is the Schwarzschild radius. 

The BHL accretion rate corresponds to the accretion rate on scales of the BH's accretion radius, $R_{\rm acc} = GM_{\rm BH}/v_{\rm wind}^2 \approx 0.3\,\rm au$. The model described above thus predicts that the accretion rate at the event horizon (and all radii interior to $\sim 10\,R_s$) is

\begin{align}
\label{eq:Mdot_EH}
\dot{M}	&\approx\left(\frac{10R_{s}}{R_{{\rm acc}}}\right)^{0.5}\dot{M}_{{\rm BHL}}=\frac{\sqrt{20}v_{{\rm wind}}}{c}\dot{M}_{{\rm BHL}} \\
    \label{eq:Mdot_EH_val}
	&\approx0.002\,\dot{M}_{{\rm BHL}},
\end{align}
i.e., a factor of $\sim500$ reduction in $\dot{M}$ relative to the BHL rate!

\begin{figure*}
    \centering
    \includegraphics[width=\textwidth]{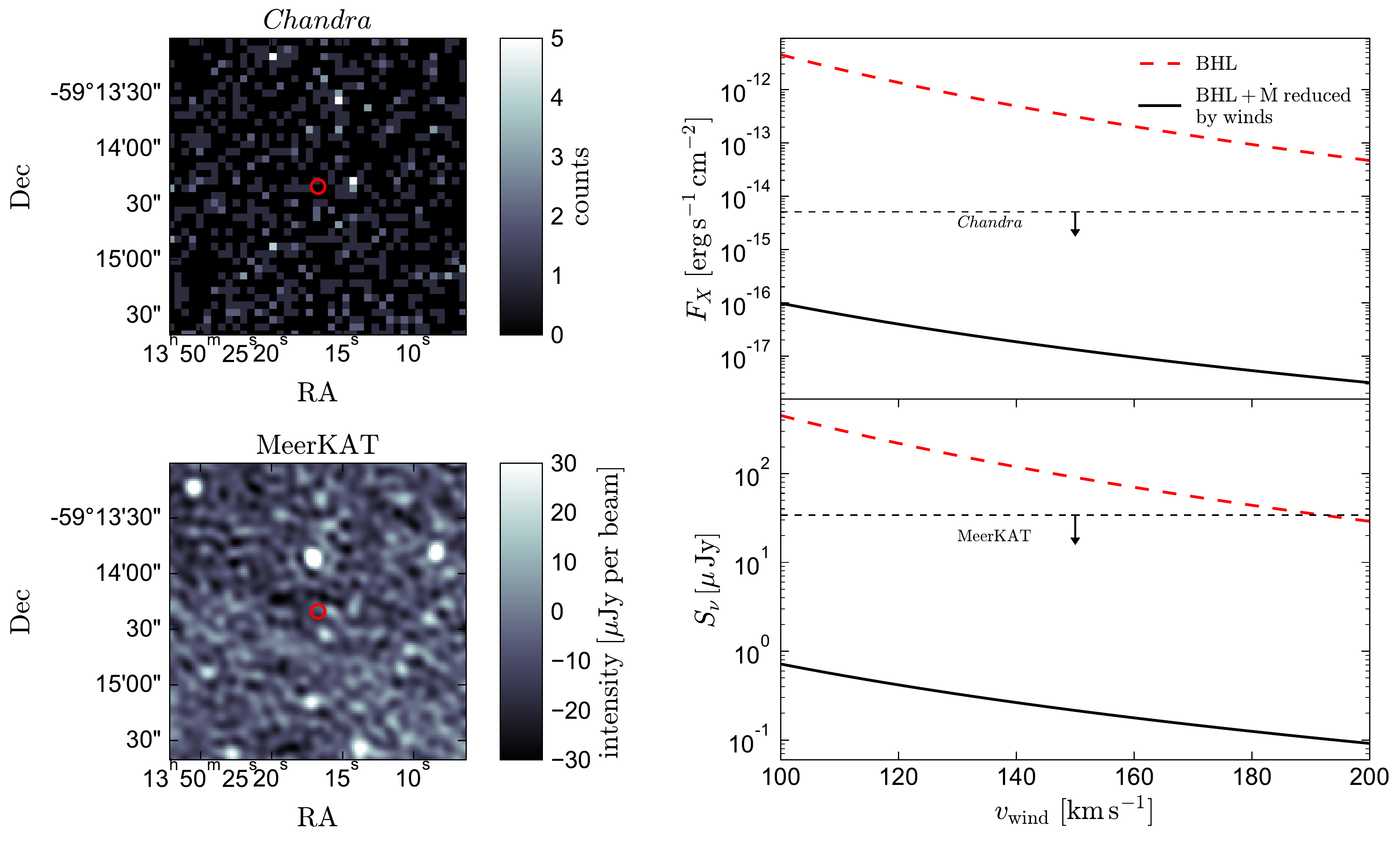}
    \caption{Left: MeerKAT and {\it Chandra} observations of Gaia BH2. Images are 2.7 arcmin wide. Red circles are centered on Gaia BH2 and have a radius of 3.5 arcsec, which is roughly the size of the MeerKAT beam and slightly larger than the {\it Chandra} PSF. The source is not detected in either image. Right: predicted X-ray and radio flux of Gaia BH2 under 2 scenarios: (1; red-dashed) the BH accretes the giant's wind at the Bondi-Hoyle-Lyttleton (BHL) rate, with the radiative efficiency calculated following models for radiatively ineffiecient accretion flows, and (2; solid black) the mass flux toward the BH decreases at smaller radii due to winds (Section~\ref{sec:wind_losses} and Equation~\ref{eq:Mdot_EH}), such that the accretion rate at the event horizon is much lower than at the Bondi radius. Dashed lines show the observed $2\sigma$ upper limits. The data are inconsistent with accretion at the BHL rate but consistent with models in which most of the would-be-accreted mass is lost to winds.}
    \label{fig:xray_radio}
\end{figure*}

A lower $\dot{M}$ also leads to a lower predicted radiative efficiency. The expected accretion rate at the horizon is $\dot{M} \approx 4\times 10^{-16} \approx 2\times 10^{-9}\dot{M}_{\rm edd}$, and the corresponding radiative efficiency from \citet{Xie2012} is predicted to be $\eta_X  \approx 7.5 \times 10^{-5}$, as opposed to  $\eta_X  \approx 0.004$ for the BHL case. Substituting this value into  Equation~\ref{eq:Fx} and finally multiplying by 0.002 (Equation~\ref{eq:Mdot_EH_val}), we obtain a predicted $F_X\approx 10^{-17}\,\rm erg\,s^{-1}\,cm^{-2}$ for Gaia BH2, which is more than 2 orders of magnitude below the {\it Chandra} limit. The corresponding radio flux is $\approx 0.2\mu$Jy, more than 2 orders of magnitude below the MeerKAT limit. The predicted X-ray and radio fluxes are shown with solid black lines in Figure~\ref{fig:xray_radio}.

To summarize: if we take the accretion rate onto the BH to be the BHL rate predicted by Equation~\ref{eq:mdot_bhl}, then the predicted radiative efficiency is $\eta_X = 0.004$, and the X-ray flux predicted by Equation~\ref{eq:Fx} is $\approx 4\times 10^{-13}\,\rm erg\,s^{-1}\,cm^{-2}$, which would have been easily detected. On the other hand, if we reduce the predicted accretion rate to account for a reduction in $\dot{M}$ between the accretion radius and the horizon (Equation~\ref{eq:Mdot_EH}), the predicted accretion rate and radiative efficiency both fall significantly, and the expected X-ray and radio fluxes would not be detectable. The nondetection of Gaia BH2 thus supports models in which winds, convection, or other processes cause a significant reduction in the horizon accretion rate relative to the BHL rate. Such models have also been supported by polarization measurements of emission from Sgr A* and M87*, which show that the accretion rate near the event horizon is 2-3 orders of magnitude less than the Bondi value \citep{Quataert2000, Kuo2014}.

\begin{figure*}
    
\end{figure*}

\section{Nature of the companion}
\label{sec:nature_of_companion}
Since we have not detected the companion directly, we can infer its nature only from (a) dynamical constraints on its mass and (b) the fact that it emits little if any detectable light. The combination of {\it Gaia} astrometry and our RVs yields a robust and precise constraint on its mass: $8.9\pm 0.3\,M_{\odot}$. This relies on astrometric constraints on the orbital inclination, which are not accessible with RVs alone.

It is instructive to disregard the astrometric inclination constraint and consider only the RVs. The orbital solution from our joint fit has a spectroscopic mass function, $f\left(M_{2}\right)=P_{{\rm orb}}K_{\star}^{3}\left(1-e^{2}\right)^{3/2}/\left(2\pi G\right)=1.33\pm0.01\,M_{\odot}$. This represents the absolute minimum companion mass compatible with the RV orbit, in the limit of an edge-on orbit and a luminous star with zero mass. Adopting $M_{\star} = 1.07\pm 0.19\,M_{\odot}$ and assuming an edge-on orbit (in conflict with the astrometric solution), this implies an minimum companion mass of $2.64\pm 0.16\,M_{\odot}$ (Figure~\ref{fig:mass_fn}). 

We now consider what kinds of companions could have the dynamically implied mass. As demonstrated in Figure~\ref{fig:seds}, a $9\,M_{\odot}$ main-sequence star would outshine the giant at all wavelengths. It would have a bolometric luminosity of $L_2 \gtrsim 5,000\,L_{\odot}$, at least 200 times brighter than the giant. An evolved  $9\,M_{\odot}$ star would be even more luminous, so any $9\,M_{\odot}$ star is ruled out by the non-detection of a second component in the UVES spectra and by the faintness of the source in the {\it Swift} UVM2 data. A $9\,M_{\odot}$ companion is also too massive to be a white dwarf or neutron star. While a small cluster of white dwarfs or neutron stars could escape detection, such a configuration would be difficult to assemble and would rapidly become unstable. We conclude that if the {\it Gaia} astrometric solution is correct, the companion must be a BH, or a close binary containing a BH.

What if the astrometric constraint on the inclination is spurious? In that case -- if the period and eccentricity from the {\it Gaia} solution are still reliable -- the companion mass could be as low as $\approx 2.6\,M_{\odot}$. This is still too massive to be a white dwarf or a neutron star, and a $2.6\,M_{\odot}$ main-sequence companion is inconsistent with the SED and spectra (3rd panel of Figure~\ref{fig:seds}). We also consider the possibility that the companion is an unresolved binary consisting of two stars with total mass $2.6\,M_{\odot}$. An equal-mass inner binary will yield the highest mass-luminosity ratio, and the bottom panel of Figure~\ref{fig:seds} shows that this scenario is also ruled out by the faintness of the source in the UV and the spectral limit in the blue optical. 

There is also an evolutionary problem for scenarios in which the companion contains one or more main-sequence stars: in all cases, the stars would be more massive than the giant itself, and should thus be more evolved than it is. But one or more red giant companions would outshine the giant in the optical.

\subsection{Could the {\it Gaia} orbit be spurious? }
Since our RV follow-up has covered only a limited fraction of Gaia BH2's orbit, it is in principle possible that the {\it Gaia} orbit is simply wrong. In practice, our follow-up RVs alone constrain the orbit fairly precisely (Figure~\ref{fig:rvonly}), excluding any orbit with $P_{\rm orb} \lesssim 1150$ days or $f(M_2)_{\rm RVs} < 1.25\,M_{\odot}$. While we do not have access to the epoch-level astrometric data on which the {\it Gaia} solution is based, it is very unlikely that the observed RVs would just happen to match the predictions of the {\it Gaia} solution if that solution were incorrect.  The {\it Gaia} orbital solution also appears unproblematic based on its goodness of fit metrics (Appendix~\ref{sec:gaia_appendix}), and the source was observed by {\it Gaia} almost 90 times over a 1000-day period, sampling most of the orbital phase (Appendix~\ref{sec:appendix_gost}). Gaia BH2 has a \texttt{AstroSpectroSB1} solution. In order to obtain an \texttt{AstroSpectroSB1} solution in {\it Gaia} DR3, a source must first have independent spectroscopic and astrometric solutions, which can be combined in a manner that improves the goodness of fit relative to the purely astrometric solution \citep{Pourbaix2022}. This means that if the solution were spurious, the spectroscopic and astrometric orbits would independently have to be spurious in compatible ways -- a scenario that is unlikely to occur for a source with such good astrometric phase overage. And finally, the minimum companion mass implied by the RVs alone -- $2.5\,M_{\odot}$ -- is still too high to be a white dwarf, or a main-sequence star that could escape spectroscopic and photometric detection. 

More broadly, the fact that the measured RVs are in good agreement with the predictions of the {\it Gaia} solution, and the {\it Gaia} fit quality metrics are unremarkable, leaves us with little reason to doubt the reliability of the {\it Gaia} solution, including the inclination constraint. In this case, the inclination is robustly constrained to $34.9\pm 0.4$ degrees, and the companion mass to $8.9\pm 0.3\,M_{\odot}$. A companion of this mass must contain at least one BH. A single BH with $M\sim 8.9\,M_{\odot}$ thus seems to be the simplest explanation, although the data do not rule out an inner binary containing 2 BHs or a BH and another compact object or low-mass star.

\subsection{Comparison to other known BHs}

\begin{figure*}
    \centering
    \includegraphics[width=\textwidth]{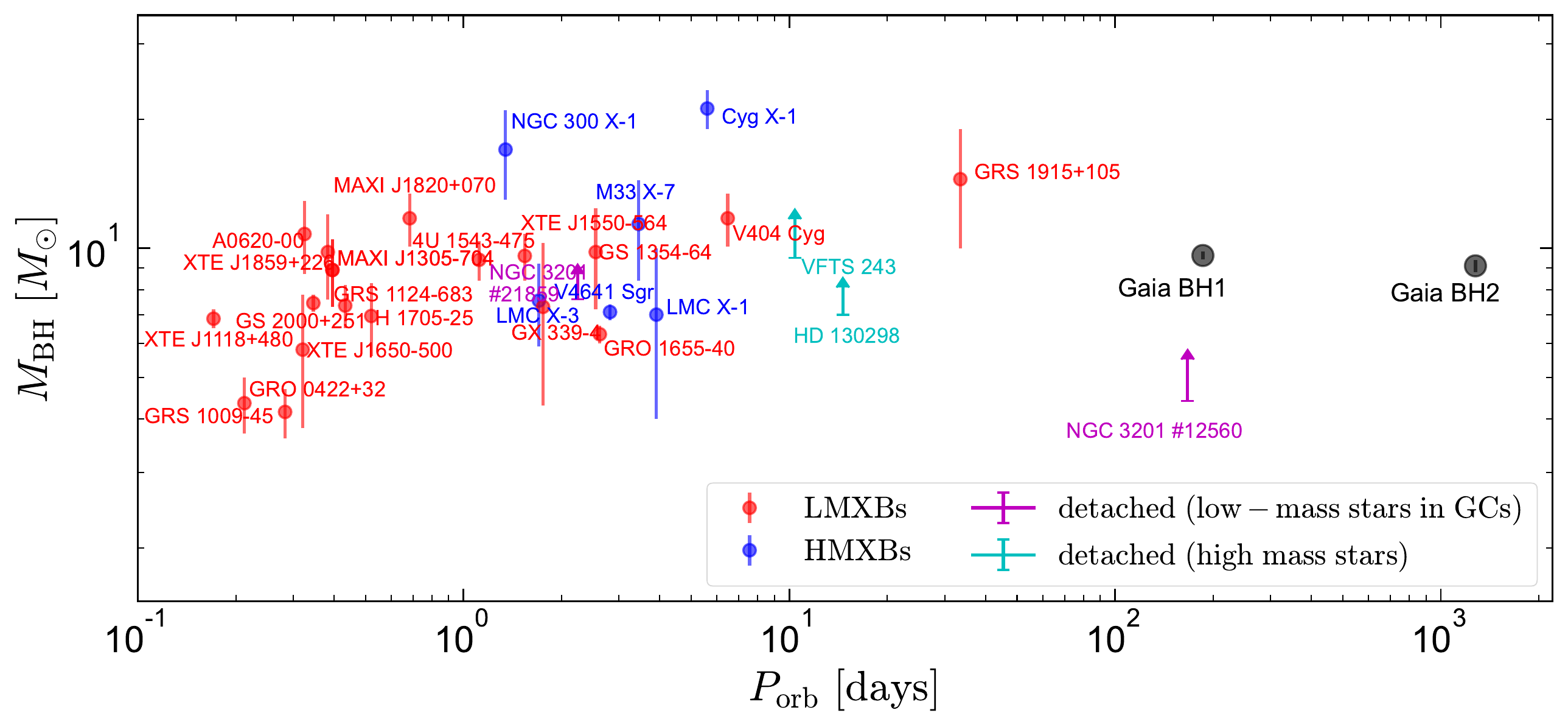}
    \caption{Comparison of Gaia BH1 and BH2 (black points) to other known BHs. Red and blue symbols correspond to accreting BHs with low- and high-mass companions. Magenta symbols show detached binaries in the globular cluster NGC 3201, and cyan points show detached binaries in which the luminous star is a high-mass ($\gtrsim 20\,M_{\odot}$) star. Gaia BH2 stands out from the rest of the population most strongly due to its orbital period, which is $7\times$ longer than that of Gaia BH1 and $\sim 1000$ times longer than the periods of typical X-ray binaries.}
    \label{fig:bh_pop}
\end{figure*}

\begin{figure*}
    \centering
    \includegraphics[width=\textwidth]{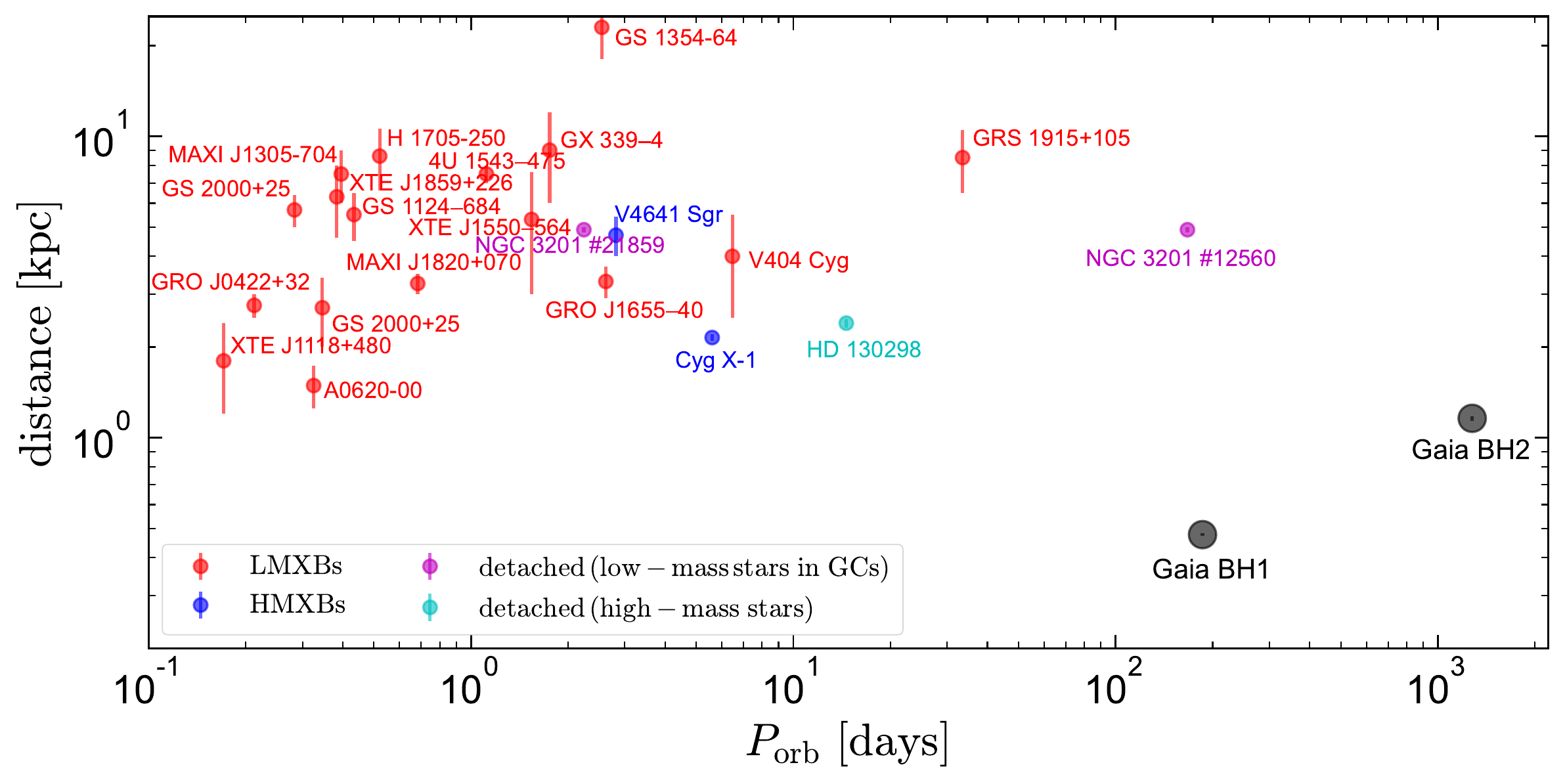}
    \caption{Comparison of Gaia BH1 and BH2 (black points) to known Galactic BHs in the plane of distance and orbital period. Color scheme is as in Figure~\ref{fig:bh_pop}. The {\it Gaia}-discovered systems are in entirely different part of this parameter space from other known BHs: they both have longer periods and are closer to Earth than any other known BHs. }
    \label{fig:bh_pop_distance_porb}
\end{figure*}

\begin{figure*}
    \centering
    \includegraphics[width=\textwidth]{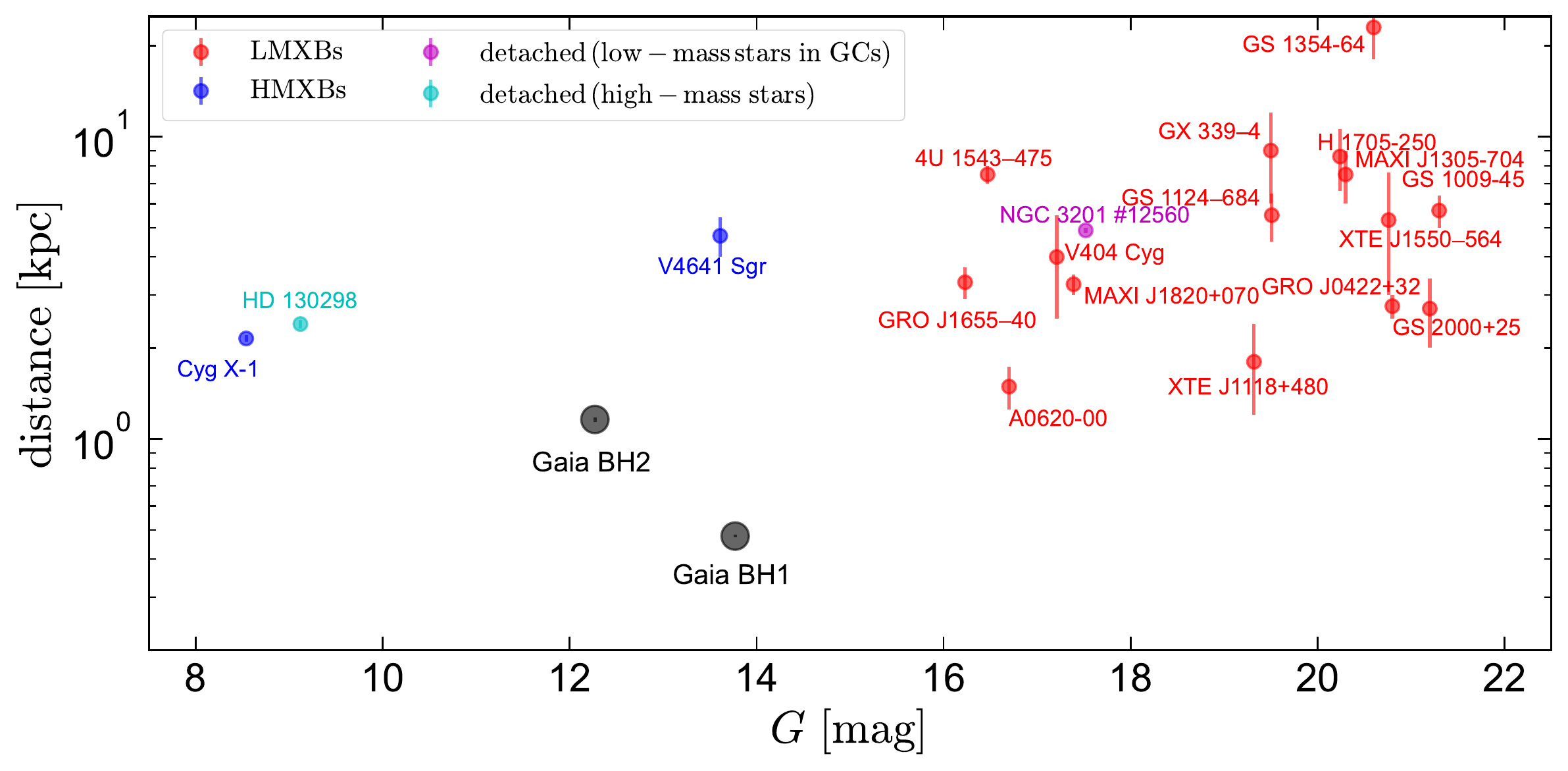}
    \caption{Comparison of Gaia BH1 and BH2 (black points) to known Galactic BHs in the plane of distance and quiescent optical magnitude. Color scheme is as in Figure~\ref{fig:bh_pop}. Gaia BH1 and BH2 are nearer than any of the other objects. They are significantly brighter than the BH LMXBs (red), but not the HMXBs, whose donors are intrinsically much more luminous.}
    \label{fig:bh_pop_distance}
\end{figure*}

In Figures~\ref{fig:bh_pop}-\ref{fig:bh_pop_distance}, we compare Gaia BH2 and BH1 to other known BHs.  Red and blue points show low- and high-mass X-ray binaries, whose parameters we take from \citet{Remillard2006} and the \texttt{BlackCAT} catalog of X-ray transients introduced by \citet{Corral-Santana2016}. 
We also show the binaries VFTS 243 (in the LMC; \citealt{Shenar2022}) and HD 130298 (in the Milky Way; \citealt{Mahy2022}), which both are single-lined binaries containing $\sim 25\,M_{\odot}$ O stars and unseen companions suspected to be BHs. In magenta, we show two binaries with suspected BH companions in the globular cluster NGC 3201, discovered with MUSE \citep[][]{Giesers2018, Giesers2019}. 

For the BHs in the Milky Way -- which includes all the LMXBs but only 2 of the HMXBs -- we additionally collected distance estimates and quiescent $G$-band magnitudes. 5 systems -- A0620-00, Cyg X-1, V4641 Sgr, GRO J1655-40, and HD 130298 -- have reasonably precise {\it Gaia} DR3 pallaxes, with \texttt{pallax\_over\_error} > 5. For these sources, we use the geometric distance estimates from  \citet{Bailer-Jones2021}, which are informed by the parallax and a Galactic model prior. For the other sources, which are fainter and/or more distant, the parallaxes are not very constraining, and so we collect distance estimates from the literature that are based on the orbital period, donor spectral type, and apparent magnitude in quiescence. Where available, we take these estimates from \citet{Jonker2004}. For the objects not included in that work, we take the estimates from individual-object papers: \citet{Casares2009} for GS 1354-64, \citet{Heida2017} for GX 339-4, \citet{Mikolajewska2022} for MAXI J1820+070, and  \citet{MataSanchez2021} for MAXI J1305-704. A few sources are too faint in the optical (in some cases due extinction) to be detected by {\it Gaia}; these are not shown in Figure~\ref{fig:bh_pop_distance}. 

Gaia BH2 has the longest orbital period of all currently known and suspected BHs. Its orbital period is a factor of seven longer than that of Gaia BH1, and $\approx 40$ times longer than that of the longest-period X-ray binary, GRS 1915+105. The masses of Gaia BH1 and BH2 are similar to the bulk of the BH X-ray binary population; only GRS 1009-45 and GRO J0422+32 host BHs of significantly lower mass.

Both Gaia BH1 and BH2 are closer, and have longer orbital periods, than any other BH binaries. They are brighter than any of the BH X-ray binaries with low-mass donors, but fainter than the HMXB Cyg X-1 (which hosts a $\sim 400,000\,L_\odot$ luminous star).
Given that their companion stars are cool and slowly rotating, their abundances and RVs can be measured with higher fidelity than in other BH companions. The two systems' close distances imply that widely separated BH binaries significantly outnumber close and accreting BH X-ray binaries. 

\section{Formation history}
\label{sec:formation_history}

As with Gaia BH1 \citep{El-Badry2023}, it is unclear whether Gaia BH2 formed from a primordial binary or via a more exotic channel. We first consider formation from a primordial binary and then discuss alternative channels in Section~\ref{sec:other_channels}.

\subsection{Natal kick constraints}
The formation of a compact object after core collapse is expected to impart a natal kick to the compact object, modifying the orbital size and eccentricity. Given the wide current orbit of Gaia BH2, it is unlikely that tides have appreciably circularized the binary following the formation of the BH. Assuming that the system formed from a pre-existing binary, its semi-major axis and eccentricity may have been changed by an asymmetric supernova kick and by instantaneous mass loss \citep[e.g. ][]{Blaauw1961}. We used \texttt{COSMIC} \citep{Breivik2020} and \texttt{emcee} \citep{emcee2013} to model the combined binary properties at core collapse and natal kick strength based on the present day orbital period and eccentricity of Gaia BH2 using the $1\sigma$ measurement errors. We followed a similar procedure to the one described in \citet{Wong2022} and \citet{El-Badry2023}. But instead of modeling the primordial binary properties, we initialized our model with the BH progenitor mass $12\,M_{\odot}$ just before core collapse, secondary mass $1.07\,M_{\odot}$, and zero eccentricity. This effectively marginalizes over all previous evolution up to the point of core collapse of Gaia BH2's progenitor. Our choice of BH progenitor mass is defined by the \citet{Fryer2012} ``delayed'' model and produces a BH with mass $9\,M_{\odot}$. The Blaauw kick associated with the instantaneous mass loss imparts an eccentricity of $\sim0.3$, so a moderate natal kick is required to produce the observed eccentricity, assuming that the orbit was previously circularized by tides. 

The strength of the natal kick is degenerate with the orbital separation at core collapse, $a_{\rm{cc}}$, which we therefore include as a parameter in our model. We used uniform priors for the orbital separation at core-collapse between $1$ and $5\,\rm{au}$, the natal kick velocity between $0$ and $100\,\rm km\,s^{-1}$, and natal kick angles that are uniform on a sphere. We ran $1024$ walkers for $50,000$ steps, thinned the chains by a factor of $10$, and retained only the last $2000$ steps to ensure a proper burn in. We find that the present-day orbital period and eccentricity prefer a core-collapse semimajor axis $a_{\rm{cc}} = 2.9\pm 0.93\,\rm{au}$ and natal kick speed $v_{\rm{kick}} = 36^{+21}_{-11}\,\rm km\,s^{-1}$.
Assuming Gaia BH2 formed as a primordial binary, this suggests that the semi-major axis was likely smaller in the past than its present-day value. The exact constraints depend on the assumed pre-supernova mass of the progenitor, which is uncertain. However, a generic result of this modeling is that the observed wide orbit and moderate eccentricity disfavor strong natal kicks to the BH. 

\subsection{BH progenitor mass}
Given the BH's mass of $\sim 9\,M_{\odot}$, its progenitor likely had an initial mass $M_{\rm init}\gtrsim 25\,M_{\odot}$ \citep[e.g.][]{Sukhbold2016, Raithel2018}. Because more massive stars are predicted to have stronger winds, the relation between initial mass and BH mass for stars of near-solar metallicity is predicted to be non-monotonic, and -- given current observational constraints -- essentially any initial mass $M_{\rm init}\gtrsim 25\,M_{\odot}$  could plausibly have produced a $9\,M_{\odot}$ BH. 

\subsection{Problems with formation through common envelope evolution}
If Gaia BH2 formed as a primordial binary with a $\gtrsim 25\,M_{\odot}$ primary and a $\sim 1 M_{\odot}$ secondary, mass transfer through Roche lobe overflow of the primary would almost certainly lead to an episode of common envelope evolution due to the extreme mass ratio. This is predicted to reduce the orbital separation by a factor of $100-1000$, depending on the structure of the primary when mass transfer began (see the discussion in \citealt{El-Badry2023}). Since Roche lobe overflow could not have begun at a separation significantly larger than 10\,au, this leads to a predicted post-common envelope separation at least a factor of 50 closer than observed. Given the system's low space velocity and modest eccentricity, we also cannot invoke a fine-tuned natal kick to the BH to widen the orbit again after a common-envelope episode. The same considerations also disfavored a common envelope for Gaia BH1, and Gaia BH2's wider orbit exacerbates the tension. 

\subsection{Formation from a star too massive to become a red supergiant}
We next consider whether Gaia BH2 could have formed from a massive star that never became a red supergiant and thus avoided mass transfer altogether. We modeled the evolution of the progenitor with single massive star evolutionary models of solar metallicity, calculated with MESA by \citet{Klencki2020}. The orbital separation at the time of the progenitor's death is unknown, since a natal kick of unknown magnitude and direction could have modified the orbit in a variety of ways. But as discussed above, the modest eccentricity suggests that the final pre-supernova separation was likely not too different from the current semi-major axis ($\sim 5$\,au). We adopt a value of 3 au. 

\begin{figure}
    \centering
    \includegraphics[width=\columnwidth]{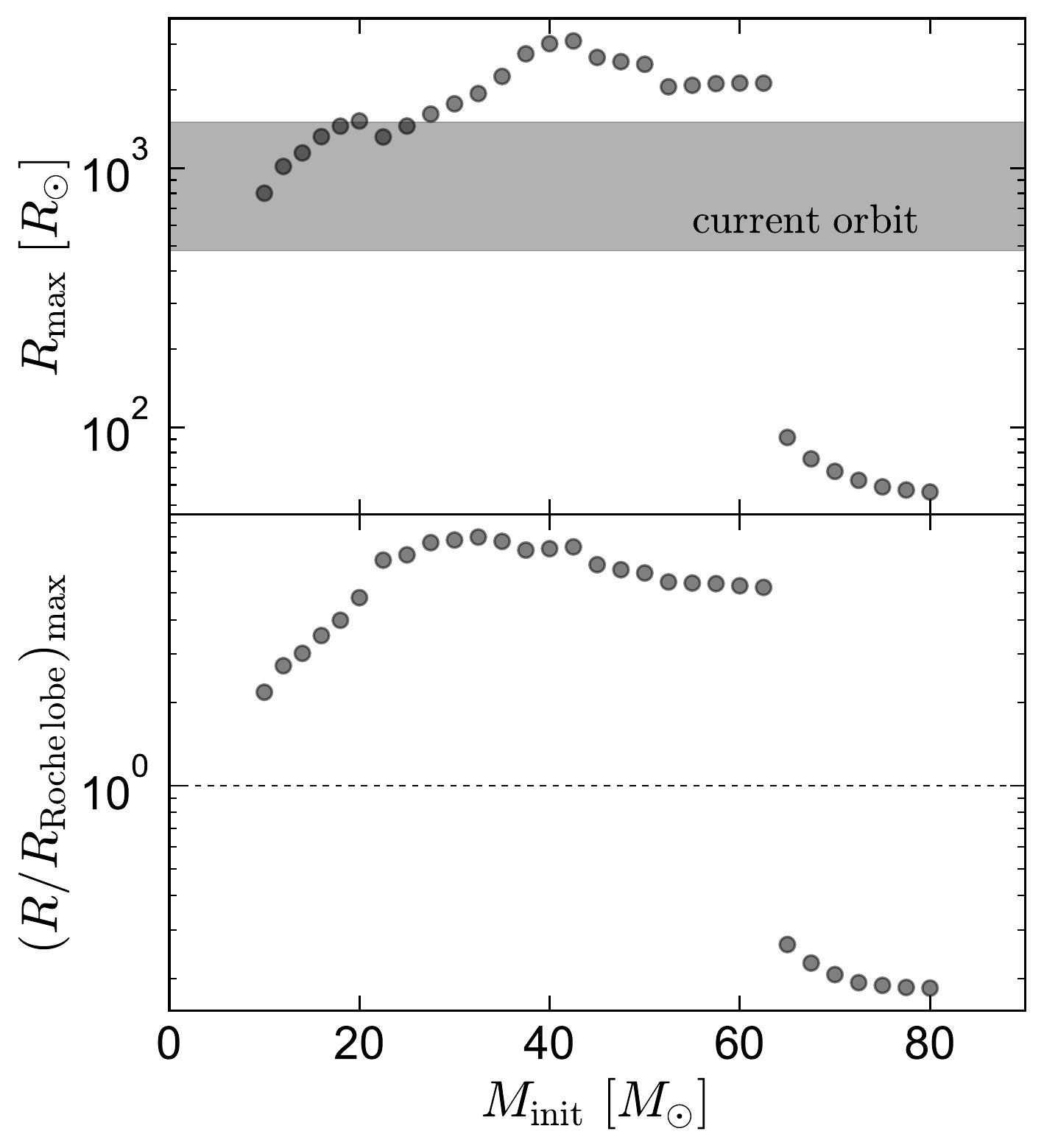}
    \caption{Predictions of single massive star evolutionary models for different initial masses of the BH progenitor. Top panel shows the maximum radius reached by the model, and gray shaded region shows the periastron-to-apastron separation of Gaia BH2's current orbit. Bottom panel shows the maximum ratio of the star's radius to the Roche lobe radius, assuming a pre-supernova separation of 3\,au and adiabatic orbital evolution due to mass loss.  For all progenitor masses below 65\,$M_{\odot}$, the model would overflow its Roche lobe in the Gaia BH2 system, presumably leading to an episode of common envelope evolution. Progenitors with initial masses $M_{\rm init}>65\,M_{\odot}$, however, lose their envelopes to strong winds at the end of their main-sequence evolution and never become red supergiants. Such models could have avoided a common envelope episode in the Gaia BH2 system. }
    \label{fig:massivestar}
\end{figure}

The massive star models undergo significant mass loss through winds. Assuming this occurs isotropically on a timescale that is long compared to the orbital period, the orbit will expand such that the product of the semi-major axis and total binary mass is conserved. This means that the orbit must have been tighter in the past in the absence of mass transfer. For each massive star model, we assume a final separation of 3\,au, and then calculate the separation at each timestep based on the mass at that time. We also calculate the Roche lobe radius, using the fitting formula from \citet{Eggleton1983} and assuming a $1\,M_{\odot}$ companion. Finally, we calculate the ratio of the massive star model's radius to the Roche lobe. We plot the maximum value of this quantity across all timesteps in Figure~\ref{fig:massivestar}. If this ratio is larger than 1, it means that the BH progenitor would have overflowed its Roche lobe at some point if it were in the Gaia BH2 system, leading to a common envelope inspiral.

All models with $M_{\rm init} \leq 65\,M_{\odot}$ would overflow their Roche lobes in the Gaia BH2 system by more than a factor of 2, meaning that they could not have avoided a common envelope. However, the most massive models behave quite differently: these models lose their hydrogen envelopes to strong winds while still on the main sequence and never become red giants \citep[see e.g.][]{Smith2008}. If the progenitor of Gaia BH2 followed such an evolutionary channel, the system could have formed as, e.g., a $70+1\,M_{\odot}$ primordial binary with separation $\sim 0.5\,\rm au$, completely avoided mass transfer through Roche lobe overflow, and died as a Wolf Rayet + $1\,M_{\odot}$ binary with separation $\sim 3$ au. The mass threshold above which stars avoid a red supergiant phase varies significantly between models, and is as low as $\sim 40\,M_{\odot}$ in some models \citep[e.g.][]{Agrawal2020, Bavera2022}.

An important caveat to these calculation is that the maximum radius of massive stars depends sensitively on parameters that are uncertain, particularly their wind mass loss rates. Some recent studies \citep[e.g.][]{Beasor2020} suggest that winds of red supergiants with $M\gtrsim 25\,M_{\odot}$ are weaker than predicted by classical models. This may imply that the mass loss rates of more massive stars are also lower, in which case envelope loss while on the main sequence might not occur. The mass threshold above which models avoid a red supergiant phase also depends significantly on assumptions about rotation \citep{Maeder1987}. We conclude that there are significant uncertainties in the evolutionary models for massive stars that avoid expansion, but it is possible that Gaia BH2 formed through such a channel.

\subsection{Other formation channels}
\label{sec:other_channels}
It is also possible that Gaia BH2 did not form from a primordial binary. One possibility is that the system formed dynamically through an exchange interaction in a dense cluster. The binary's thin disk-like Galactic orbit precludes formation in a classic globular cluster, but the system may have formed in an open cluster that has since dissolved. This scenario does not leave any smoking gun signatures and is thus difficult to test definitively for any single binary. N-body simulations of binaries in clusters \citep[e.g.][]{Kremer2018} that are tailored to systems like Gaia BH1 and BH2 are required to determine whether such dynamical exchange+ejection events in clusters are common enough to explain the observed rate of wide BH + normal star binaries. {\it If} the observed $\alpha-$enhancement of the giant is a result of pollution from the BH progenitor (see Section~\ref{sec:abund_interp}), this would disfavor a dynamical formation channel. 

It is also possible that Gaia BH2 formed as a triple, with the red giant an outer tertiary to an inner binary containing two massive stars. Interactions between the two stars could have prevented either one from becoming a red supergiant. This scenario is discussed further by \citet{El-Badry2023} for Gaia BH1; a similar scenario could also work for Gaia BH2. Indeed, a triple scenario likely requires less fine-tuning for Gaia BH2 than for BH1, because the wider orbit makes it possible to accommodate a wider range of inner binaries without the system becoming unstable. As with Gaia BH1, precision RV monitoring of the luminous star may eventually detect or rule out a dark inner binary, which would cause a short-period, low-amplitude modulation of the overall RV curve.

\subsection{Future evolution}
Over the next $\approx 100\,\rm Myr$, the luminous star is expected to continue ascending the red giant branch, until it ignites core helium burning. The MIST models shown in Figure~\ref{fig:mosaic} predict that if it evolved in isolation, the giant would reach a radius of about $170\,R_{\odot}$ just before the helium flash. However, in the current binary orbit, the giant would overflow its Roche lobe at periastron when it reached $R\approx 110\,\rm R_{\odot}$. The orbit will likely circularize as the giant expands, and there will likely be a brief period of mass transfer through stable Roche lobe overflow near the tip of the giant branch, when the system will be observable as a symbiotic X-ray binary.
After helium ignition, the giant will contract and the binary will appear as a wide BH binary with a red clump or hot subdwarf star. Another episode of stable Roche lobe overflow is possible after core helium burning, depending on how much of the star's envelope mass remains. Either way, the binary is expected to terminate its evolution as a wide BH + white dwarf binary, probably with a low-eccentricity orbit.

\section{An early assessment of the BH population revealed by \textit{Gaia} DR3 }
\label{sec:future}
\subsection{Completeness of follow-up}

We identified both Gaia BH1 and BH2 as BH candidates using simple astrometric cuts designed to find massive and dark companions to low-mass stars (see Appendix E of \citealt{El-Badry2023}). We have now followed-up all of the 6  candidates those cuts yielded. Two have been validated, and the remaining four turned out to have spurious {\it Gaia} orbital solutions. 

A broader search for compact object companions among the {\it Gaia} DR3 astrometric solutions was carried out by \citet{Shahaf2023}. Gaia BH2 was excluded from their sample both because the luminous source is not on the main sequence and because its orbital period is longer than 1000 days. Besides Gaia BH1, \citet{Shahaf2023} identified an additional seven BH candidates, whose {\it Gaia} orbital solutions appeared to imply a dark companion with $M_2 > 2\,M_{\odot}$. We are carrying out RV follow-up of these sources, which we will describe in detail elsewhere. Only for one of them, {\it Gaia} DR3 6328149636482597888, do the RVs inspire confidence in the {\it Gaia} solution. However, that source is a low-metallicity halo star, and so the masses of both the luminous star and the unseen companion are lower than inferred when the system is interpreted with solar-metallicity models (see \citealt{El-Badry2023}). The best-fit companion mass is $M_2 \approx 2.2\,M_{\odot}$. This is consistent with a variety of astrophysical scenarios; we defer more detailed analysis of this source to future work. 

In summary, we suspect that Gaia BH1 and BH2 are the {\it only} BHs with DR3 astrometric orbital solutions, at least among sources with implied companion masses $M_2\gtrsim 2.5\,M_{\odot}$. There are, however, a few dozen candidates with $1 \lesssim M_2 \lesssim 2\,M_{\odot}$ -- most readily interpreted as neutron stars and white dwarfs -- still to be followed-up.

\subsection{Is there a BH -- neutron star mass gap?}
It is conspicuous that there are now two vetted 
{\it Gaia} compact object binary solutions with $8 < M_2/M_{\odot} < 10$, a few dozen candidates with  $1 \lesssim M_2 \lesssim 2.5\,M_{\odot}$, and {\it no} reliable solutions with  $3 < M_2/M_{\odot} < 8$. The implied BH/NS companion mass distribution is reminiscent of the mass gap observed for BHs in X-ray binaries \citep[e.g.][]{Kreidberg2012}, though the mean BH mass inferred for X-ray binaries is somewhat lower ($\sim 7\,M_{\odot}$). While {\it Gaia} is somewhat more sensitive to high-mass companions, which produce a larger photocenter wobble at fixed orbital period, both Gaia BH1 and BH2 would plausibly have been detectable if the BH had a mass of, say, $5\,M_{\odot}$. The data are thus already in tension with a scenario in which low-mass BHs significantly outnumber $\sim$$10\,M_{\odot}$ BHs, at least among BHs with stellar companions in wide orbits. 
The fact that \citet{Shahaf2023} identified several dozen wide neutron star candidates with dynamically-implied masses of $1-2\,M_{\odot}$ also suggests that {\it Gaia} would be sensitive to lower-mass BHs, but further follow-up is required to vet the neutron star candidates.

The next decade will likely see the discovery of both more wide BH + normal star astrometric binaries and isolated single BHs discovered via astrometric microlensing \citep[e.g.][]{Sahu2022, Lam2022, Mroz2022}. Comparison of the mass distributions of BHs within these populations may shed light on the formation pathways of the BH binaries.

\subsection{Orbital period distribution}
\label{sec:porb_dist}

\begin{figure}
    \centering
    \includegraphics[width=\columnwidth]{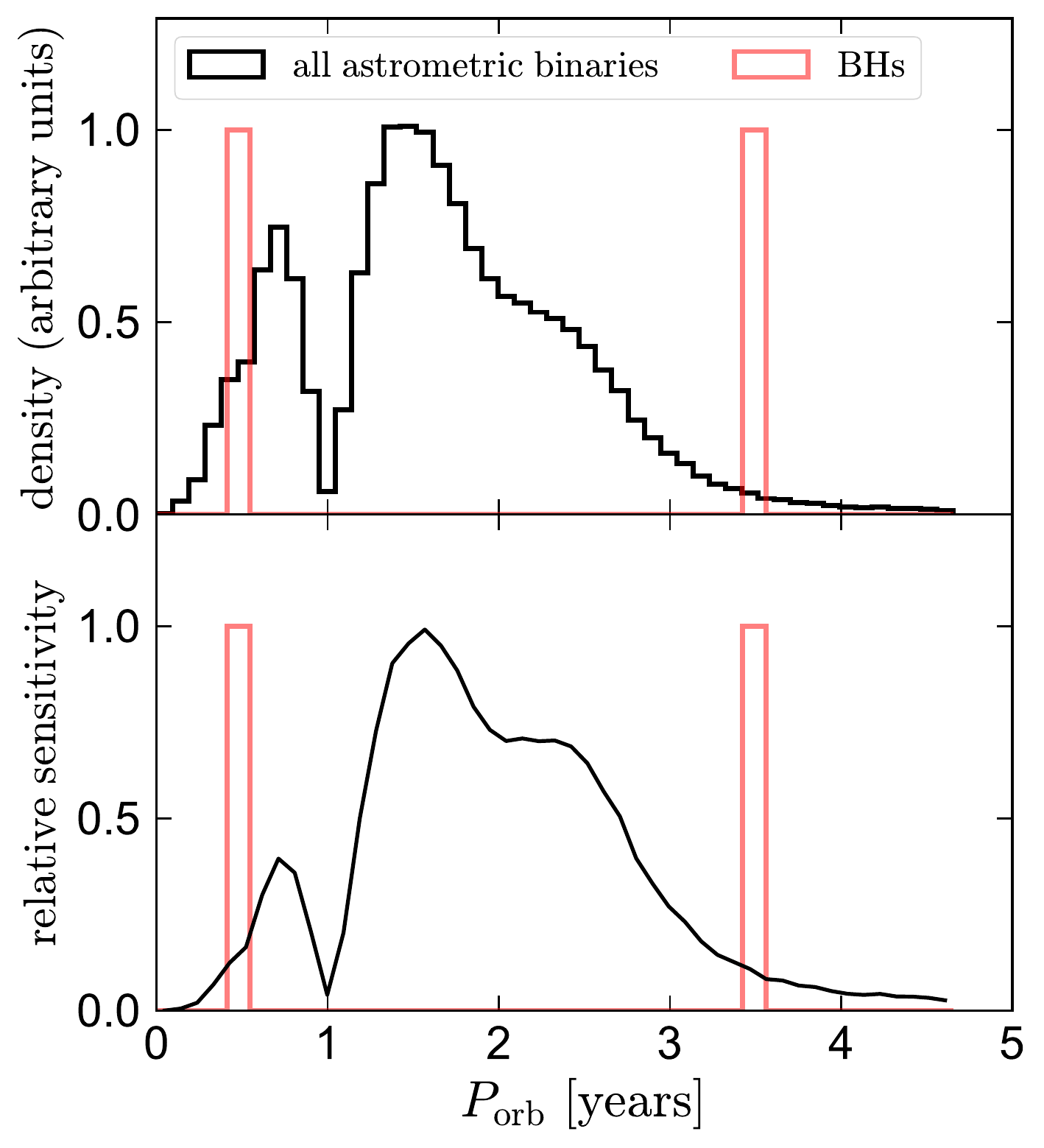}
    \caption{Orbital period distribution of all astrometric binaries in {\it Gaia} DR3 (black) and Gaia BH1 and BH2 (red). In the bottom panel, we divide the observed period distribution of all astrometric binaries by the expected intrinsic binary period distribution for solar-type stars. This provides an approximate sensitivity function of {\it Gaia} DR3 as a function of orbital period, marginalized over other observables. The two BHs have periods at the long- and short-period edges of the full observed period distribution, where DR3's sensitivity is $\sim 10\times$ lower than at its peak at 1.5 years. This is suggestive of a bimodal intrinsic period distribution, with fewer BHs at periods of 1-3 years.}
    \label{fig:porb_distribution}
\end{figure}

In Figure~\ref{fig:porb_distribution}, we compare the orbital periods of Gaia BH2 and BH1 to the period distribution of all $\approx 186,000$ astrometric binaries published in DR3. Most of these systems are solar-type binaries, whose intrinsic period distribution is well-approximated by a lognormal that peaks at a period of $10^5$ days with a dispersion of 2.3 dex \citep{Raghavan2010}. To estimate the relative sensitivity of {\it Gaia} DR3 to binaries at different periods, we divide the observed period distribution by this lognormal. The latter decreases toward longer periods (for linear period intervals), so this rescaling gives more weight to binaries with long periods. 

Figure~\ref{fig:porb_distribution} shows that {\it Gaia} DR3 was most sensitive to binaries with orbital periods between 1.25 and 2.5 years, since shorter periods produce smaller astrometric wobbles at fixed mass, and longer periods could not be fully sampled during the $\sim 1000$ day observing baseline for data published in DR3. The strongly reduced sensitivity to periods of one year, due to confusion of orbital and parallax motion, is also evident. The fact that the two BHs discovered from {\it Gaia} DR3 have orbital periods on the long- and short-period edges of the observational sensitivity curve hints that the BH companion period distribution may be bimodal, with a valley at 1-3 year periods. The strength of the conclusions we can draw is of course limited by the fact that we have detected only two BHs. 
We can nevertheless draw some conclusions.  If the intrinsic period distribution of BH binaries were {\it flat} in linear orbital period, 80\% the detected binaries would have periods between 1 and 3 years. There is thus only a $0.2^2 = 4\%$ chance of detecting two BHs outside this range and none inside it if the underlying BH period distribution is flat. 

As discussed by \citet{El-Badry2023}, {\it Gaia} DR4 and DR5 will likely enable the discovery of dozens of BH + normal star wide binaries, both because they will be based on a longer observing baseline than DR3 and because epoch-level astrometric data will be published for all sources, without stringent cuts on signal-to-noise ratio. DR4 will be based on $\sim$2000 days of observations, and DR5 will likely be based on $\sim$4000 days of observations. {\it Gaia} will thus eventually enable the discovery and characterization of BH binaries with periods up to $\sim 15$ years (and perhaps longer if partial orbits can be fit uniquely; e.g., \citealt{Andrews2021}). The intrinsic period distribution of BH + normal star binaries is quite uncertain, but some models predict that it will rise steeply at $P_{\rm orb} \gtrsim 10$ years, where the BH binaries will have formed without the BH progenitor ever having overflowed its Roche lobe \citep[e.g.][]{Breivik2017, Chawla2022}. This prediction depends sensitively on the strength of BH natal kicks, which can easily unbind the widest binaries. In any case, the empirical period and mass distributions of BH binaries will become much better sampled in future {\it Gaia} releases. 

\section{Conclusions}
\label{sec:concl}
We have carried out follow-up observations of a newly identified black hole (BH) + luminous star binary discovered via {\it Gaia} astrometry and spectroscopy. The system, which we call Gaia BH2, consists of an apparently unremarkable lower red giant branch star in a 3.5-year orbit with an unseen massive companion. Basic properties of the system are summarized in Table~\ref{tab:system}. Our main conclusions are as follows:

\begin{enumerate}
    \item {\it The luminous source}: the luminous source is a bright ($G=12.3$), nearby ($d=1.16\,\rm kpc$) star on the lower red giant branch, just below the red clump (Figure~\ref{fig:mosaic}). From evolutionary models, we infer a mass $M_\star = 1.07\pm 0.19\,M_{\odot}$. From a high-resolution spectrum (Figure~\ref{fig:modelspec}), we measured moderately subsolar metallicity ($\rm [Fe/H] = -0.22$) but strong enhancement of $\alpha-$elements ($\rm [\alpha/Fe] = +0.26$). The detailed abundance pattern is otherwise unremarkable (Table~\ref{tab:bacchus}). Because the system's orbit is wide, the $\alpha-$enhancement is most likely primordial, but it is also possible that it is a result of pollution from low-velocity ejecta during a failed supernova.
    \item {\it Orbit and companion mass}: our RV follow-up over a 7-month period validates the {\it Gaia} orbital solution (Figure~\ref{fig:rvfig}) and yields improved constraints on the companion mass (Figure~\ref{fig:corner_plot_comparison}). The orbit has a 1277-day period -- longer by a factor of 7 than any other known BH binary -- and is moderately eccentric, with $e=0.52$. The inclination is $\approx 35$ deg; i.e., relatively face-on. The dynamically implied mass of the companion is $M_2 = 8.9\pm 0.3\,M_{\odot}$. This constraint depends only weakly on the assumed mass of the luminous star, but significantly on the inclination (Figure~\ref{fig:mass_fn}). 
    \item {\it Limits on luminous companions}: the SED from the UV to the near-IR (Figure~\ref{fig:seds}) and high-resolution spectrum (Figures~\ref{fig:modelspec} and~\ref{fig:galah}) are both well fitted by single-star models. A luminous companion contributing more than $\approx 3\%$ of the light at 4000\,\AA\,\, is ruled out spectroscopically. The giant is cool ($T_{\rm eff} = 4600\,\rm K)$, and so the fact that the source is faint in the UV rules out any scenario in which the companion is one or more main-sequence stars (Figure~\ref{fig:seds}). These limits rule out any plausible luminous stellar companion if the {\it Gaia} solution is correct. Even if the inclination inferred from the {\it Gaia} orbit were spurious -- in which case the RVs would allow a companion mass as low as $2.6\,M_{\odot}$ -- a stellar companion would be detectable. 
    \item {\it Nature of the companion}: the dynamically implied mass of the companion -- $8.9\pm 0.3\,M_{\odot}$ -- is too large to be a neutron star or white dwarf, or any plausible combination of white dwarfs, neutron stars, and faint stars. The simplest explanation is a single BH. As with Gaia BH1, it is possible that the companion is an unresolved binary containing at least one BH. 
    \item {\it Formation history}: the current orbit of Gaia BH2, with a periastron distance of $2.4\,\rm au$, is too close to accommodate BH progenitors with initial masses $20 \lesssim M_{\rm init}/M_{\odot} \lesssim 65$, which are predicted to become red supergiants with radii larger than the current orbit (Figure~\ref{fig:massivestar}). One possibility is that the BH progenitor had a sufficiently high mass ($M_{\rm init}\gtrsim 65\,M_{\odot}$, though the precise value varies between models) to lose its envelope while on the main sequence, avoiding a red supergiant phase.  It is also possible that the system was formed dynamically through an exchange interaction in a dense cluster, or through triple evolution.
    \item {\it X-ray and radio observations}: we observed Gaia BH2 with both X-ray and radio facilities but did not detect it (Figures~\ref{fig:xray_radio}). Expressed in terms of the Bondi-Hoyle-Lyttleton (BHL) accretion rate, the non-detection implies a radiative efficiency $\lesssim 10^{-4}$.   The non-detection is consistent with model predictions for radiatively-inefficient accretion flows in which most of the inflowing material is lost to winds or accretion is suppressed by convection, such that only a small fraction of the BHL accretion rate actually reaches the event horizon. For typical radiative efficiencies, the non-detection is inconsistent with models in which a mass flux equal to the BHL rate reaches the event horizon.
    
    \item {\it The {\it Gaia} BHs in context}: {\it Gaia} DR3 has thus far resulted in the identification of two widely separated BH + normal star binaries, whose properties are quite different from previously known BH binaries (Figures~\ref{fig:bh_pop}-\ref{fig:bh_pop_distance}). Both have longer periods and are closer to Earth than any previously known BHs. This suggests that widely-separated BH binaries significantly outnumber close systems with ongoing mass transfer. Both have masses between 8 and 10\,$M_{\odot}$. There do not appear to be any BHs with astrometric solutions published in DR3 that have masses between 3 and 8 $M_{\odot}$. The two BHs have periods at the short- and long-period edges of the {\it Gaia} sensitivity curve (Figure~\ref{fig:porb_distribution}), perhaps hinting at a bimodal intrinsic period distribution for BH binaries. 
\end{enumerate}

\section*{Acknowledgements}
We thank Frédéric Arenou, Jim Fuller, and Andy Gould for useful discussions, and  Maren Hempel, Paul Eigenthaler, and Régis Lachaume for carrying out the FEROS observations. We are grateful to the ESO, MeerKAT, {\it Chandra}, and {\it Swift} directorial offices and support staff for prompt assistance with DDT observations.

HWR acknowledges the European Research Council for the ERC Advanced Grant [101054731].
This research used pystrometry, an open source Python package for astrometry timeseries analysis \citep[][]{Sahlmann2019}. This work used Astropy,\footnote{http://www.astropy.org} a community-developed core Python package and an ecosystem of tools and resources for astronomy \citep{AstropyCollaboration2022}, and the CIAO software \citep{Fruscione2006} provided by the Chandra X-ray Center (CXC). We acknowledge the use of public data from the {\it Swift} data archive.

This project was developed in part at the Gaia Fête, held at the Flatiron Institute's Center for Computational Astrophysics in June 2022, and in part at the Gaia Hike, held at the University of British Columbia in June 2022. 
This work has made use of data from the European Space Agency (ESA) mission
{\it Gaia} (\url{https://www.cosmos.esa.int/gaia}), processed by the {\it Gaia}
Data Processing and Analysis Consortium (DPAC,
\url{https://www.cosmos.esa.int/web/gaia/dpac/consortium}). Funding for the DPAC
has been provided by national institutions, in particular the institutions
participating in the {\it Gaia} Multilateral Agreement.

The MeerKAT telescope is operated by the South African Radio Astronomy Observatory, which is a facility of the National Research Foundation, an agency of the Department of Science and Innovation. The scientific results reported in this article are based in part on observations made by the Chandra X-ray Observatory, which is operated by the Smithsonian Astrophysical Observatory for and on behalf of the National Aeronautics Space Administration under contract NAS8-03060.

\section*{Data Availability}
Data used in this study are available upon request from the corresponding author. 



\bibliographystyle{mnras}

\appendix

\section{{\it Gaia} observations}
\label{sec:appendix_gost}

\begin{figure*}
    \centering
    \includegraphics[width=\textwidth]{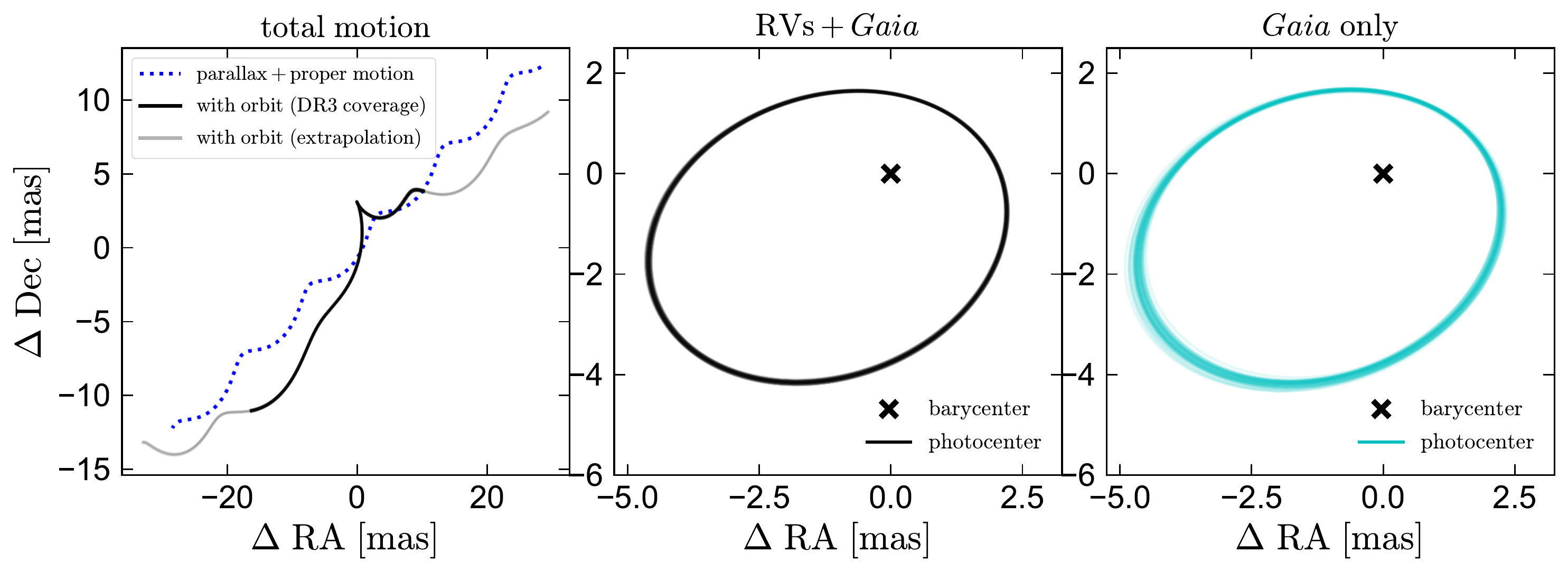}
    \caption{Left: plane-of-the-sky photocenter motion as inferred from our joint RV + {\it Gaia} fit. Blue line shows the motion due to parallax and proper motion alone. Black and grey lines show the full motion, including wobble due to the star's orbit. Black line shows the time period covered by {\it Gaia} DR3 astrometry; gray line extrapolates beyond that period. Middle and right panels show constraints on the orbital motion alone, with parallax and proper motion subtracted. Right panel shows the constraints from the {\it Gaia} data alone, while middle panel shows constraints from the joint fit.}
    \label{fig:full_motion}
\end{figure*}

\subsection{Predicted astrometry}
Figure~\ref{fig:full_motion} shows the plane-of-the-sky motion of the photocenter of Gaia BH2 as it would be observed by {\it Gaia}. In the left panel, the dashed blue line shows the motion expected due to parallax and proper motion alone, while the black and gray lines show the combined effects of orbital motion, parallax, and proper motion. Lines are colored darker in the time interval covered by {\it Gaia} observations included in DR3, between July 2014 and May 2017. The interplay between parallax and proper motion leads to rather complex motion on the plane of the sky, which -- given the limited time baseline of DR3 -- does not yet appear periodic. Nonetheless, the {\it Gaia} data alone constrain the orbit fairly precisely. 

In the right panel, individual cyan lines show draws from the posterior of the fit to the pure {\it Gaia} solution, with proper motion and parallax motion removed. Although there is noticeable dispersion in these predictions owing to observational uncertainties, the size and shape of the orbital ellipse are well-determined. The middle panel shows draws from the posterior of the joint fit of the {\it Gaia} data and RVs. As expected, the resulting orbital ellipse is consistent with the {\it Gaia}-only fit, but its parameters are more tightly constrained.   

\subsection{Observation times}
The \texttt{AstroSpectroSB1} solution for Gaia BH2 is based on both astrometric and RV measurements. Single-epoch astrometric and RV data is not published in DR3, but some insight into when a source was observed can be provided by the {\it Gaia} observation scheduling tool (GOST).\footnote{https://gaia.esac.esa.int/gost/} Given a source position, GOST provides a list of observation times when the scanning law predicts that a source will transit across the {\it Gaia} focal plane. For Gaia BH2, GOST predicts 97 CCDs transits during the observation window that contributed to DR3, distributed over 27 visibility periods (i.e., groups of observations separated by at least 4 days). Meanwhile, the \texttt{gaia\_source} table reports \texttt{matched\_transits = 93} and \texttt{visibility\_periods\_used = 26}. This suggests that a majority of the planned transits indeed contributed to the DR3 solution, but a few were omitted, as is common. Fewer transits contributed RVs, with  \texttt{rv\_nb\_transits} = 32 and \texttt{rv\_visibility\_periods\_used} = 12. This is also expected, since only a fraction of focal plane transits cover the RVS spectrometer \citep{GaiaCollaboration2016}. 

\begin{figure}
    \centering
    \includegraphics[width=\columnwidth]{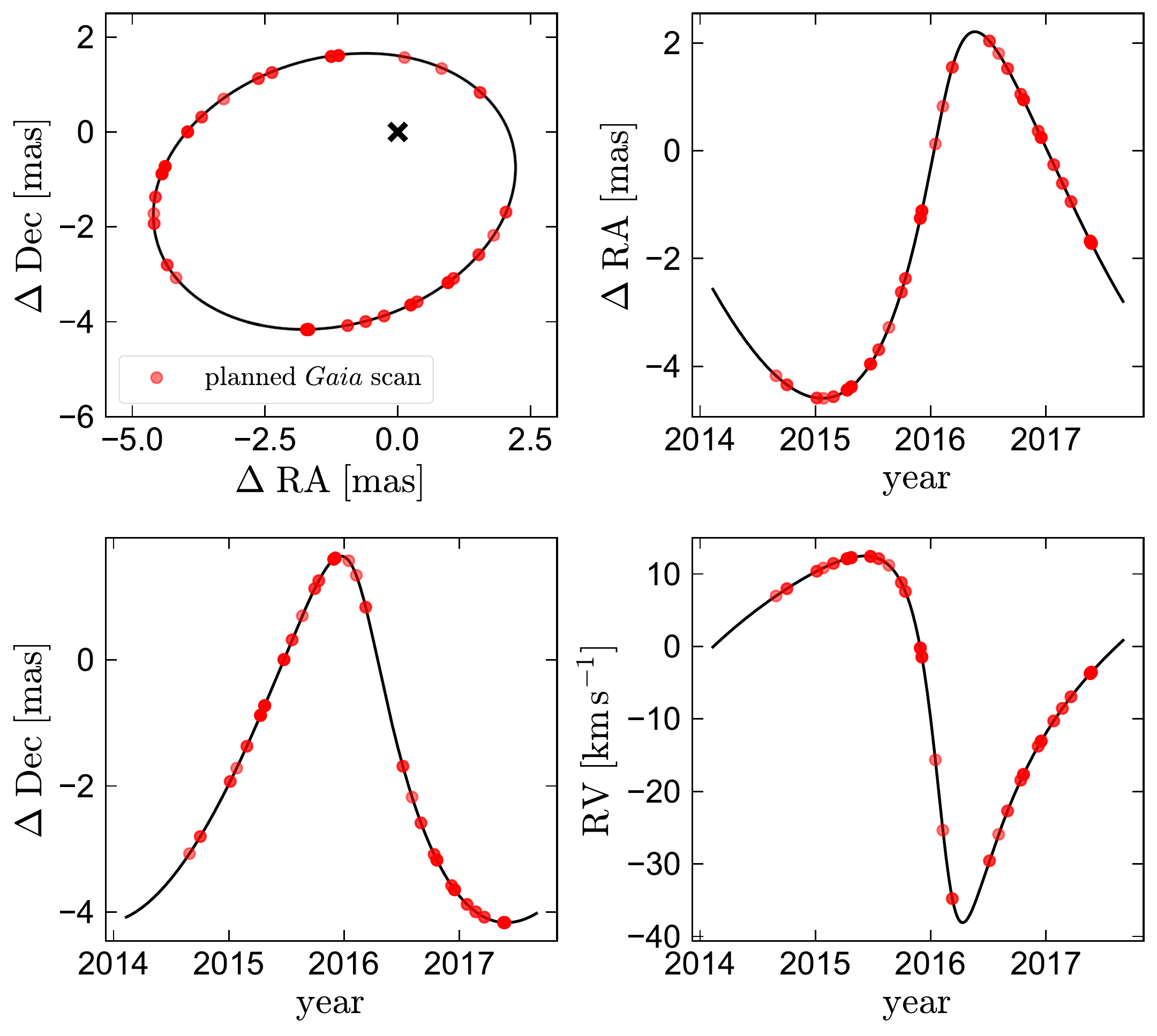}
    \caption{Predicted observation times of Gaia BH2 from the {\it Gaia} observation scheduling tool (GOST). Black line shows the best-fit orbit from our combined fit. Red points show the predicted photocenter positions at the times when GOST predicts {\it Gaia} would have observed the source. Note that we do not have access to the actual measured $\Delta\,\rm RA$, $\Delta\,\rm Dec$ and $\rm RV$ values; only to the predicted scan times. Further caveats are discussed in the text.}
    \label{fig:gost}
\end{figure}

In Figure~\ref{fig:gost}, we show the {\it predicted} astrometric positions and RVs at observation times predicted by GOST. It is important to note that (a) epoch-level data is not published in DR3, so there is no guarantee that the actual observed positions and RVs match the model predictions, (b) we have not simulated noise, (c) actual {\it Gaia} astrometric measurements are 1D, with the scan angle varying from visit to visit, and (d) while a majority (93/97) of the planned transits actually contributed data to the astrometric solution, only about one third (32/93) contributed an RV. 

We can nevertheless draw several conclusions from Figure~\ref{fig:gost}. The predicted astrometric measurement times span 1000 days -- 79 \% of the binary's orbital period -- and cover most of the predicted plane-of-the sky motion. The typical actually-realized along-scan uncertainty for each astrometric measurement for a source with $G=12.3$ is 0.2 mas \citep{Lindegren2021}, which is comparable to the size of the red points in Figure~\ref{fig:gost}. The expected per-epoch RV uncertainty for a source of Gaia BH2's apparent magnitude and spectral type is also about 1.5 $\rm km\,s^{-1}$ (see e.g. \citealt{El-Badry2019twin}, their Figure F1), also comparable to the size of the points in the lower right panel. The reported \texttt{rv\_amplitude\_robust} = $37\,\rm km\,s^{-1}$ (representing the total range in measured epoch RVs, after clipping of outliers) suggests that RVs were measured over most of the orbit's dynamic range in RV, though the two transits predicted near the periastron passage in 2016 likely did not result in an RV measurement. 

We conclude that the {\it Gaia} measurements likely are adequate to constrain the orbit robustly, though it would obviously be preferable to have access to the actual epoch-level data. 

\section{Gaia goodness of fit}
\label{sec:gaia_appendix}

\begin{figure}
    \centering
    \includegraphics[width=\columnwidth]{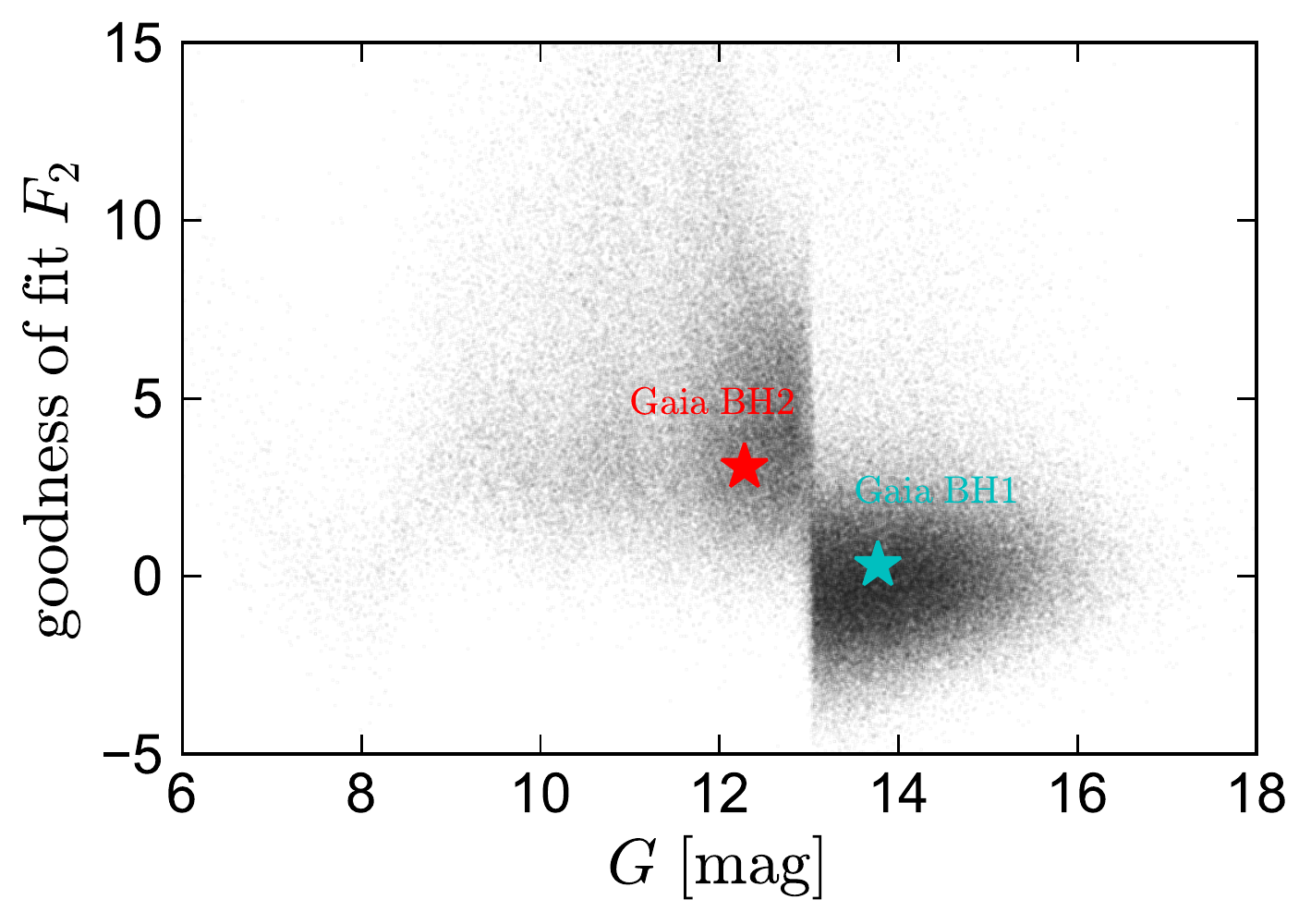}
    \caption{Reported \texttt{goodness\_of\_fit} ($F_2$; Equation~\ref{eq:F2}) for all {\it Gaia} astrometric binaries, with Gaia BH1 and BH2 highlighted. Large $F_2$ values indicate a poor formal fit. There is a discontinuity in the typical $F_2$ values at $G=13$ owing to a quirk of the {\it Gaia} data processing: brighter sources typically have larger $F_2$ values because the uncertainties in their individual-epoch astrometric data are underestimated. Both Gaia BH1 and BH2 have typical $F_2$ values for their apparent magnitude, indicative of an unremarkable astrometric solution.  }
    \label{fig:gof}
\end{figure}

One of the quality diagnostics calculated for {\it Gaia} orbital solutions is the \texttt{goodness\_of\_fit} estimator, $F_2$, which is defined  \citep[e.g.][]{Halbwachs2022} as
\begin{equation}
    \label{eq:F2}
    F_{2}=\sqrt{\frac{9\nu}{2}}\left[\left(\frac{\chi^{2}}{\nu}\right)^{1/3}+\frac{2}{9\nu}-1\right].
\end{equation}
Here $\nu$ is the number of degrees of freedom and $\chi^2$ is the sum of the uncertainty-normalized squared residuals between model and data. When there is good agreement between model and data and the uncertainties are reliably estimated, one expects $F_2$ to be normally distributed; i.e., $F_2 \sim \mathcal{N}(0,1)$. Values of $F_2 \gg 1$ thus indicate a formally poor fit. Gaia BH2 has $F_2 = 3.07$, raising potential concern about the quality of the {\it Gaia} solution. 

In Figure~\ref{fig:gof}, we compare the $F_2$ values of Gaia BH2 and BH1 to those of all astrometric binary solutions published in DR3 (including both \texttt{Orbital} and \texttt{AstroSpectroSB1} solutions). This comparison reveals that Gaia BH2 actually has a somewhat lower $F_2$ value (indicative of a better fit) than typical sources at its apparent magnitude: almost all sources with $G<13$ have $F_2 > 1$. This most likely is a result of the change in {\it Gaia} window class at $G=13$ \citep{Rowell2021}: epoch astrometry for sources with $G<13$ is derived from a 2D window fitted with a point spread function, while epoch astrometry for fainter sources is calculated from 1D data. This leads to discontinuities in many astrometric quantities at $G=13$ \citep[e.g.][]{El-Badry2021_wbs, Lindegren2021b}. 

The fact that Gaia BH2 has $F_2 > 1$ does  not necessarily imply that the uncertainties on the \texttt{AstroSpectroSB1} solution -- which also enter into our joint fit -- are underestimated. In calculating the astrometric solution, the uncertainties of the epoch-level data are inflated by multiplying by a constant, $c=\sqrt{\chi^{2}\nu^{-1}\left(1-\frac{2}{9\nu}\right)^{-3}}$, after which $F_2=0$  \citep{Halbwachs2022, Pourbaix2022}.

\section{TESS light curves}
\label{sec:tess}
\begin{figure*}
    \centering
    \includegraphics[width=\textwidth]{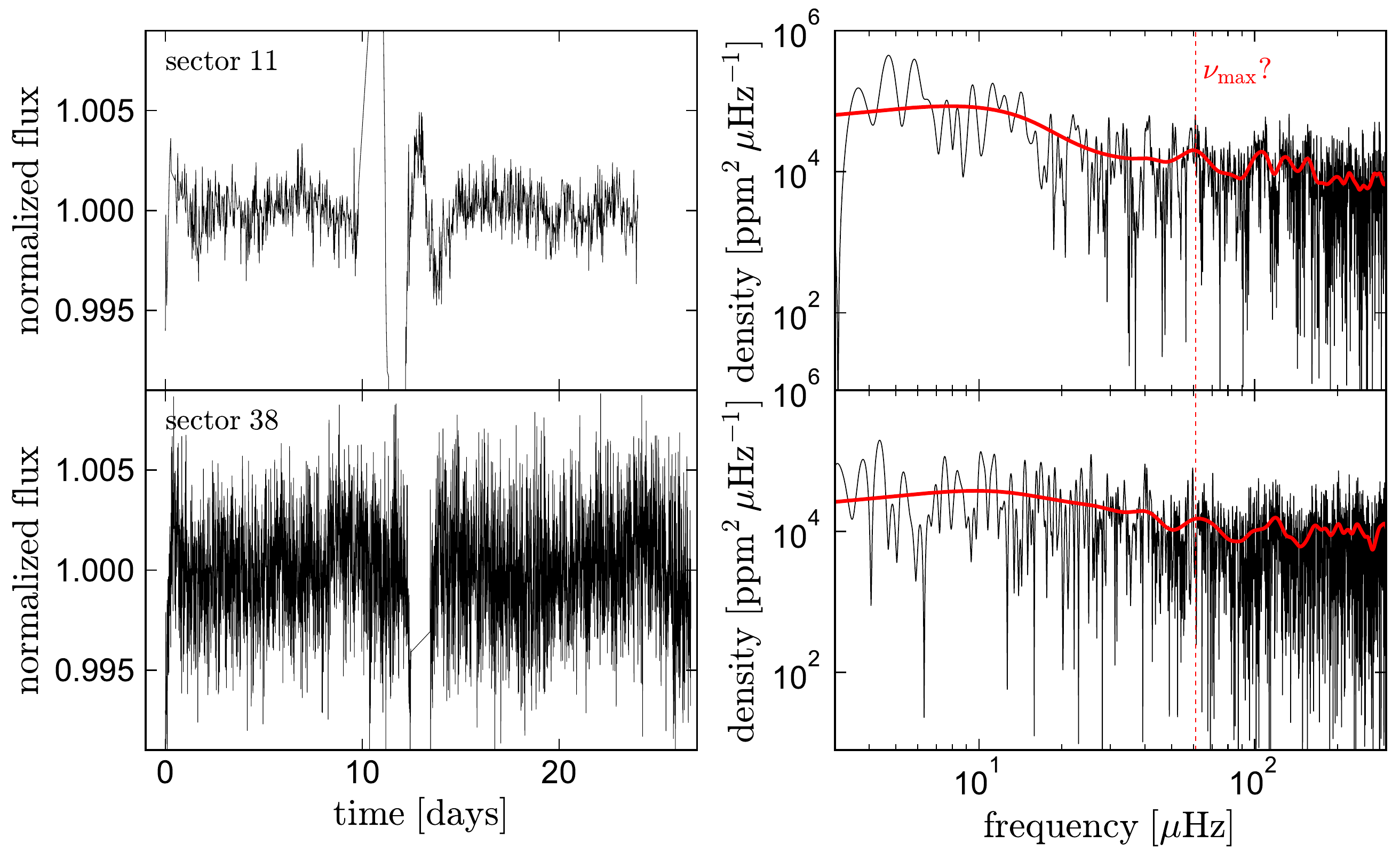}
    \caption{{\it TESS} data for Gaia BH2. Left panels show normalized light curves from the full frame images, with an exposure time of 1800s for sector 11 and 600s for sector 38. Right panels show power spectra, and red lines show power spectra smoothed with a Gaussian kernel with $\sigma = 5\,\mu \rm Hz$. Vertical dashed line shows a tentative detection of $\nu_{\rm max}$. }
    \label{fig:tess}
\end{figure*}

The \texttt{eleanor} light curves of Gaia BH2 (TIC 207885877) are shown in Figure~\ref{fig:tess}. We extract these using a $15\times 15$ pixel target pixel file, with the background estimated in a $31\times 31$ pixel region. We use the systematics-corrected flux in the default aperture chosen by \texttt{eleanor}; i.e., the aperture with the lowest combined differential photometric precision (CDPP) on one-hour timescales. The expected flux contamination from background stars in the aperture is of order 10\%. 

The source is clearly not strongly variable on short timescales. Much of the apparent variability is noise; some variability on longer timescales may be the result of imperfect background subtraction and removal of systematics. The light curve from sector 38 is noisier than the one from sector 11, mainly because sector 38 used $3\times$ shorter exposure times.  

Power spectra for both sectors are shown in the right panels of Figure~\ref{fig:tess}. The red line shows the result of smoothing the power spectra with a Gaussian kernel with $\sigma= 5\,\mu \rm Hz$. Several potentially significant peaks are evident in the smoothed power spectra. We tentatively identify the peak with $\nu = 61\,\mu \rm Hz$ as the global asteroseismic parameter $\nu_{\rm max}$, since it is the strongest feature evident in data from both sectors. We emphasize that the identification is tentative since the time baseline and signal-to-noise ratio of the {\it TESS} data is still insufficient to unambiguously constrain $\nu_{\rm max}$ in a source of this apparent magnitude \citep[e.g.][]{Stello2022}. The implied surface gravity is $\log g\approx\log g_{\odot}+\log\left(\nu_{{\rm max}}/\nu_{{\rm max,\odot}}\right)+\frac{1}{2}\log\left(T_{{\rm eff}}/T_{{\rm eff},\odot}\right)\approx2.68$, where $g$ is in cgs units and $\log g_{\odot}\approx 4.43$, $\nu_{\rm max,\odot}=3090\,\rm \mu Hz$, and $T_{\rm eff,\odot}=5777\,\rm K$.

\bsp	
\label{lastpage}
\end{document}